\pdfoutput=1
\RequirePackage[T1]{fontenc}
\documentclass[12pt]{article}

\usepackage[height=8.85in,width=6.45in]{geometry}

\usepackage[utf8]{inputenc}
\usepackage{amsmath}
\usepackage{amssymb}
\usepackage{mathtools}
\numberwithin{equation}{section}
\usepackage{slashed}
\usepackage{braket}
\usepackage{enumitem}
\usepackage[svgnames,psnames]{xcolor}
\usepackage[colorlinks,citecolor=DarkGreen,linkcolor=FireBrick,urlcolor=FireBrick,linktocpage,unicode,psdextra]{hyperref}
\usepackage{cite}
\usepackage{graphicx}
\usepackage{tikz}
\usepackage{tikz-cd}
\newcommand{\tikzmark}[1]{\tikz[remember picture,overlay]\node (#1){};}
\tikzset{>=stealth}
\usepackage[bottom]{footmisc}

\usepackage{times}
\usepackage{courier}
\usepackage{bm}
\usepackage{subfig}

\usepackage{xcolor}
\usepackage{mdframed}

\renewenvironment{figure}[1][]{
  \begin{originalfigure}[#1]
    \begin{mdframed}[linecolor=black!0,backgroundcolor=black!1]
}{
    \end{mdframed}
  \end{originalfigure}
}

\usepackage{ascmac}
\usepackage{dashbox}
\usepackage{xcolor}
\usepackage{colortbl}
\usepackage{arydshln}
\definecolor{lightyellow}{rgb}{1.0, 0.95, 0.7}
\definecolor{blue}{rgb}{0.0, 0.4, 1.0}
\definecolor{Blue}{rgb}{0,0,1}
\definecolor{darkgreen}{rgb}{0.,0.6,0.}
\newcommand*{\red}[1]{\textcolor{red}{#1}}
\newcommand*{\Blue}[1]{\textcolor{Blue}{#1}}

\newcommand*{\green}[1]{\textcolor{darkgreen}{#1}}
\newcommand*{\black}[1]{\textcolor{black}{#1}}

\definecolor{colorA}{rgb}{1,0,0}
\definecolor{colorB}{rgb}{0,0.3,1}
\definecolor{colorC}{rgb}{0.9,0.8,0.2}
\definecolor{colorD}{rgb}{0,0.65,0}
\definecolor{lesslightgray}{rgb}{0.5,0.5,0.5}

\let\tilde\widetilde

\usepackage{colortbl}
\definecolor{lightyellow}{rgb}{1.0, 0.95, 0.7}
\definecolor{lightblue}{rgb}{0.7, 0.9, 1.0}
\definecolor{lightpink}{rgb}{1.0, 0.85, 0.95}
\definecolor{lightgreen}{rgb}{0.7, 1.0, 0.4}

\def\Nequals#1{$\mathcal{N}{=}#1$}
\def\bR{\mathbb{R}}
\def\bQ{\mathbb{Q}}
\def\bZ{\mathbb{Z}}

\def\fP{\mathfrak{P}}

\def\U{\mathrm{U}}
\def\SU{\mathrm{SU}}
\def\SO{\mathrm{SO}}
\def\USp{\mathrm{USp}}
\def\u{\mathfrak{u}}
\def\su{\mathfrak{su}}
\def\so{\mathfrak{so}}
\def\usp{\mathfrak{usp}}
\def\Spin{\mathrm{Spin}}
\def\SL{\mathrm{SL}}
\def\boo{0.0}
\def\xlattice#1#2#3{
\begin{tikzpicture}[scale=.5]
\filldraw[color=black!5!white](-.5,-.5) rectangle (1.5,1.5);
\draw[->] (-1,0) -- (2,0);
\draw[->] (0,-1) -- (0,2);
\foreach \x in {0,1} {
	\foreach \y in {0,1}{
		\pgfmathsetmacro\a{mod(#1 * \x - #2 * \y,2)}
		\ifx\a\boo
			\filldraw[color=#3] (\x,\y) circle (.5em);
		\else
			\filldraw[fill=white,draw=gray] (\x,\y) circle (.5em);
		\fi
	}
}
\end{tikzpicture}
}

\begin{document}

\begin{titlepage}

\begin{flushright}
\end{flushright}

\vskip 3cm

\begin{center}

{\Large \bfseries Matching higher symmetries\\[.8em]
across Intriligator-Seiberg duality}

\vskip 1cm
Yasunori Lee$^1$, Kantaro Ohmori$^2$, and Yuji Tachikawa$^1$
\vskip 1cm

\begin{tabular}{ll}
$^1$ & Kavli Institute for the Physics and Mathematics of the Universe (WPI), \\
& University of Tokyo,  Kashiwa, Chiba 277-8583, Japan\\
$^2$ & Department of Physics, Faculty of Science, \\
& University of Tokyo, Bunkyo, Tokyo 113-0033, Japan
\end{tabular}

\vskip 1cm

\end{center}

\noindent
We study higher symmetries and anomalies of 4d  $\so(2n_c)$ gauge theory with $2n_f$ flavors.
We find that they depend on the parity of $n_c$ and $n_f$,
the global form of the gauge group, and the discrete theta angle.
The contribution from the fermions plays a central role in our analysis.
Furthermore, our conclusion applies to \Nequals1 supersymmetric cases as well, and
we see that higher symmetries and anomalies match across the Intriligator-Seiberg duality between
$\so(2n_c)\leftrightarrow\so(2n_f-2n_c+4)$.

\end{titlepage}

\setcounter{tocdepth}{2}
\tableofcontents

\newpage

\section{Introduction and summary}
\label{sec:introduction}
Our understanding of the concept of symmetries in quantum field theories (QFT) has been greatly improved in the last several years.
We now have the concept of $p$-form symmetries acting on \mbox{$p$-dimensional} operators \cite{Gaiotto:2014kfa}.
This concept  gives a point of view which unifies both
ordinary symmetries acting on point operators for $p=0$
and center symmetries of gauge theories acting on Wilson line operators for $p=1$.
In addition, the 't Hooft magnetic flux \cite{tHooft:1979rtg} can now be thought of as a background gauge field for the 1-form center symmetry.
It is also realized more recently that 0-form symmetries and 1-form symmetries can not only coexist in a direct product but also mix in a more intricate manner.
They can have mixed anomalies between them,
or they can also combine nontrivially to form a symmetry structure called 2-groups.\footnote{%
The first appearance of 2-groups in string theory is in the Green-Schwarz mechanism.
Namely, that the gauge-invariant field strength of the $B$-field is $H=dB+CS(\omega)+\cdots$ where $\omega$ is the affine connection means that the $\U(1)$ 1-form symmetry (for which the $B$-field is the gauge field) 
and the diffeomorphism form a nontrivial 2-group extension.
This point was discussed in a series of papers by Urs Schreiber and his collaborators, see
e.g.~\cite{Baez:2005sn,Sati:2008eg,Sati:2009ic,Fiorenza:2010mh,Fiorenza:2012tb}.
Note that the 2-group in this case is a gauge symmetry.
The significance of 2-groups as a \emph{global} symmetry structure in field theory was recognized much later in \cite{Sharpe:2015mja,Cordova:2018cvg,Benini:2018reh}.
}

In this paper, we study these issues in the case of 4d $\so$ quantum chromodynamics (QCD),
i.e.~$\so(N_c)$ gauge theories with 
$N_f$ flavors of fermion fields in the vector representation.
We assume that the fermions are massless, unless otherwise explicitly stated. 

\paragraph{Symmetries of $\so$ QCD:}
Let us start by quickly recalling the 0-form and 1-form symmetries of the $\so$ QCD.
As for the 0-form symmetry, 
we focus our attention on the $\su(N_f)$ symmetry acting on $N_f$ flavors of matter fields 
in the vector representation.
We will not consider other discrete symmetries in this paper for brevity.

As for the 1-form symmetry, we first need to recall that 
the theory comes in three versions, $\Spin$, $\SO_+$ and $\SO_-$,
distinguished firstly by the global form of the gauge group ($\Spin$ vs.~$\SO$)
and further by the choice of the discrete theta angle ($\SO_+$ vs.~$\SO_-$) \cite{Aharony:2013hda}.\footnote{%
This is when the theories are considered on spin manifolds.
For non-spin manifolds, a further distinction needs to be made \cite{Ang:2019txy}.
For simplicity, we only consider spin manifolds in this paper.
}
They also differ by the nontrivial line operator they possess: 
the $\Spin$ theory has the Wilson line $W$ in the spinor representation,
the $\SO_+$ theory has the 't Hooft line $H$ which is mutually non-local with respect to $W$,
and the $\SO_-$ theory has the dyonic line $D=WH$. 
Furthermore, these line operators are charged under corresponding $\bZ_2$ 1-form symmetries,
which we respectively call electric, magnetic and dyonic 1-form symmetries.

The main question is then how the $\su(N_f)$ 0-form symmetry and the $\bZ_2$ 1-form symmetry are related.\footnote{%
A partial answer was given in \cite{Hsin:2020nts}, but the contribution from fermions was not taken into account in that reference.
Our conclusion is consistent with theirs when the fermion contributions vanish, e.g.~when the fermions can be made massive preserving the flavor symmetry.
}
We concentrate on the case when $N_c$ and $N_f$ are both even: $N_c=2n_c$ and $N_f=2n_f$.
We introduce three possible behaviors, which we call \emph{\bfseries none}, \emph{\bfseries extension}, and \emph{\bfseries anomaly}.

\bigskip

\begin{itemize}[leftmargin=1em]
\item The case \emph{\bfseries none}. 
The $0$-form symmetry and the $1$-form symmetry stay separate without an anomaly.

\item The case \emph{\bfseries extension}.
Take, for example, the $\Spin(2n_c)$ gauge theory with $2n_f$ flavors.
When $n_c$ is odd,
two copies of the Wilson line $W$ in the spinor representation
form a Wilson line in the vector representation.
This can be screened by a dynamical fermion, which is why $W^2=1$ 
as far as the 1-form symmetry charge is concerned.
Now let us recall that this dynamical fermion transforms nontrivially under $-1\in \SU(2n_f)$.
Therefore, when we further take the flavor symmetry into account, $W^2$ is still nontrivial.

As was discussed in \cite{Hsin:2020nts},
this means that the $\bZ_2$ 1-form symmetry extends the $\SU(2n_f)/\bZ_2$ 0-form symmetry in a nontrivial manner, forming a 2-group $H$ fitting in the sequence\begin{equation}
	0
	\longrightarrow \bZ_2[1]
	\longrightarrow H
	\longrightarrow \SU(2n_f)/\bZ_2
	\longrightarrow 0,\label{2-group}
\end{equation}
whose extension class is specified by \begin{equation}
\beta a_2 \in H^3(B(\SU(2n_f)/\bZ_2);\bZ_2).\label{postnikov}
\end{equation}
Here, $\bZ_2[1]$ stands for $\bZ_2$ regarded as a 1-form symmetry,
$\SU(2n_f)/\bZ_2$ is the quotient by the subgroup $\{\pm1\}$,
$a_2$ is (the representative cocycle of) the obstruction class controlling whether the $\SU(2n_f)/\bZ_2$ bundle
lifts to $\SU(2n_f)$,
and $\beta$ is the Bockstein homomorphism.\footnote{%
\label{foot:bock}
For a $\bZ_2$-valued cocycle $a\in Z^2(X,\bZ_2)$, its Bockstein is defined as follows.
We first construct the $\bZ_4$-valued lift $\overline{a}$ of $a$ by sending $\{0,1\}$ to $\{0,1\}\subset \{0,1,2,3\}$.
Let us now consider $\delta\overline{a}$.
By construction it is 0 mod 2, and therefore $\tfrac12\delta\overline{a}$ is a well-defined $\bZ_2$-valued cocycle,
which is defined to be $\beta a$.
When $\beta a$ is zero as a cohomology class, there is a $\bZ_2$-valued cochain $b$ such that $\beta a=\delta b$. 
This is equivalent to the fact that the $\bZ_4$-valued cochain $\overline{a}':=\overline{a}+2b$ is a $\bZ_4$-valued cocycle.
In this manner we found that $[\beta a]=0$ means that $a$ can be lifted to a $\bZ_4$-valued cocycle.
We now lift this $\bZ_4$-valued  cocycle to a $\bZ_8$-valued cochain $\overline{\overline{a}}$.
In this case $\tfrac14\delta\overline{\overline{a}}$ is a well-defined $\bZ_2$-valued cocycle, 
which is defined to be $\beta_2 a$. 
This $\beta_2$ is known as a higher Bockstein operation, which we will need to use later in the paper.
When $\beta_2 a$ is zero as a cohomology class, we can lift $a$ to a $\bZ_8$-valued cocycle.
We can then lift it to a $\bZ_{16}$-valued cochain and define $\beta_3 a$, ad infinitum.
}
The background field for such a 2-group is given by the background gauge field for $\SU(2n_f)/\bZ_2$ together with a $\bZ_2$-valued degree-2 cochain $E$ satisfying\footnote{%
In this equation, $a_2$ and $\beta a_2$ need to be interpreted as cochains rather than cohomology classes.
More generally, cochains and cohomology classes will not be carefully distinguished explicitly in this paper. 
Hopefully this bad practice would not cause too much confusions.
} \begin{equation}
\delta E=\beta a_2.\label{delta}
\end{equation}
A consequence of this nontrivial extension is that the theory cannot be coupled to a general $SU(2n_f)/\mathbb{Z}_2$ background without introducing a nontrivial $E$ background.
In short, $SU(2n_f)/\mathbb{Z}_2$ is not a subgroup but a quotient group of the whole symmetry 2-group, and thus gauging $SU(2n_f)/\mathbb{Z}_2$ alone without gauging $\mathbb{Z}_2[1]$ part does not make sense.

\item The case \emph{\bfseries anomaly}.
The $\SO_+(2n_c)$ gauge theory is obtained by gauging the $\bZ_2$ 1-form symmetry of the $\Spin(2n_c)$ gauge theory \cite{Kapustin:2014gua}.
The gauging of the $\bZ_2$ 1-form symmetry whose background field is $E$ is done by 
introducing another $\bZ_2$-valued degree-2 closed cochain $B$, adding the interaction \begin{equation}
2\pi i \cdot \frac12\int BE,
\label{coupling}
\end{equation}
and summing over all possible $E$. 
In this particular case, the background field $E$ is the second Stiefel-Whitney class $w_2$ of the $\SO(2n_c)$ gauge bundle,
and summing over them gives the $\SO_+$ gauge theory.

As we will see, the contribution to the anomalies from the fermions significantly complicates the analysis.
Neglecting this contribution, we see that 
the coupling \eqref{coupling} is not closed due to \eqref{delta}, and has the variation 
\begin{equation}
2\pi i \cdot \frac12 \int B\beta a_2.\label{mixed}
\end{equation}
This means that the $\bZ_2$ 1-form symmetry of the $\SO(2n_c)$ gauge theory and the $\SU(2n_f)/\bZ_2$ 0-form flavor symmetry remains a direct product, but with a mixed anomaly given by \eqref{mixed}.
\end{itemize}

We will carefully analyze how the $\bZ_2$ 1-form symmetry and the $\SU(2n_f)/\bZ_2$ 0-form symmetry are combined in the rest of the paper.
The novelty in our paper over the analysis in \cite{Hsin:2020nts} is that we take fermionic contributions into account.
The derivation will be detailed in the following, and here we simply summarize the result  in Table~\ref{table:main}.

\begin{table}
\centering
\begin{tabular}{r@{\,}l|ccc}
$(n_c,$&$n_f)$ & $\Spin$ & $\SO_+$ & $\SO_-$\\
\hline
\vphantom{$\Bigm($}
(even,&even) & no\tikzmark{A}ne & no\tikzmark{B}ne & no\tikzmark{C}ne \\
\vphantom{$\Bigm($}
(odd,&even) & exte\tikzmark{P}nsion & ano\tikzmark{Q}maly & exte\tikzmark{R}nsion \\
\vphantom{$\Bigm($}
(even,&odd) & ano\tikzmark{S}maly & exte\tikzmark{T}nsion & exte\tikzmark{U}nsion \\
\vphantom{$\Bigm($}
(odd,&odd) & exte\tikzmark{L}nsion & exte\tikzmark{M}nsion & ano\tikzmark{N}maly 
\end{tabular}
\caption{How the $\bZ_2$ 1-form symmetry and the $\SU(2n_f)/\bZ_2$ 0-form symmetry are combined
in massless $\so(2n_c)$ QCD.
`none' implies that they remain a direct product without mixed anomaly;
`anomaly' means that they remain a direct product but with mixed anomaly;
and `extension' is when they combine into a nontrivial 2-group. 
The orange lines show how the duality of Intriligator and Seiberg acts on this set of theories.
\label{table:main}}

\tikz[overlay,remember picture]{%
\draw[<->,bend left,color=Orange,line width=1.5] (A.north) to (C.north);
\draw[<->,color=Orange,line width=1.5] (B.north west) .. controls +(75:0.5) ..  (B.north east);
\draw[<->,bend left,color=Orange,line width=1.5] (P.north) to (R.north);
\draw[<->,color=Orange,line width=1.5] (Q.north west) .. controls +(75:0.5) ..  (Q.north east);
\draw[<->,color=Orange,line width=1.5] (S.south) to ([yshift=.2em]N.north);
\draw[<->,color=Orange,line width=1.5] ([yshift=+.2em]T.south) to ([yshift=0em]M.north);
\draw[<->,color=Orange,line width=1.5] (U.south) to ([yshift=.2em]L.north);
}
\end{table}

So far we assumed that the fermions are massless.
It is also useful to see what happens when the fermions are massive.
When we give equal masses to all $N_f=2n_f$ fermions, the flavor symmetry is reduced from $\SU(2n_f)/\bZ_2$ to $\SO(2n_f)/\bZ_2$.
The crucial simplification is that $\beta a_2$ appearing in the anomaly or the extension becomes cohomologically trivial when $n_f$ is odd, 
because $a_2$ now lifts to a $\bZ_4$ class controlling whether an $\SO(2n_f)/\bZ_2$ bundle lifts to a $\Spin(2n_f)$ bundle.
Stated differently, the contributions from the fermions vanish since the fermions can be made massive,
so that then the analysis of \cite{Hsin:2020nts} applies.
The results are shown in Table~\ref{table:sub},
which is significantly simpler than the behavior in Table~\ref{table:main}.

\begin{table}
\centering
\begin{tabular}{r@{\,}l|ccc}
$(n_c,$&$n_f)$ & $\Spin$ & $\SO_+$ & $\SO_-$\\
\hline
\vphantom{$\Bigm($}
(even,&even) & no\tikzmark{AA}ne & no\tikzmark{BB}ne & no\tikzmark{CC}ne \\
\vphantom{$\Bigm($}
(odd,&even) & exte\tikzmark{PP}nsion & ano\tikzmark{QQ}maly & exte\tikzmark{RR}nsion \\
\vphantom{$\Bigm($}
(even,&odd) & no\tikzmark{SS}ne & no\tikzmark{TT}ne & no\tikzmark{UU}ne \\
\vphantom{$\Bigm($}
(odd,&odd) & no\tikzmark{LL}ne & no\tikzmark{MM}ne & no\tikzmark{NN}ne 
\end{tabular}
\caption{How the $\bZ_2$ 1-form symmetry and the $\SO(2n_f)/\bZ_2$ 0-form symmetry are combined
in massive $\so(2n_c)$ QCD.
Our conventions follow that of Table~\ref{table:main}.
\label{table:sub}}

\tikz[overlay,remember picture]{%
\draw[<->,bend left,color=Orange,line width=1.5] (AA.north) to (CC.north);
\draw[<->,color=Orange,line width=1.5] (BB.north west) .. controls +(75:0.5) ..  (BB.north east);
\draw[<->,bend left,color=Orange,line width=1.5] (PP.north) to (RR.north);
\draw[<->,color=Orange,line width=1.5] (QQ.north west) .. controls +(75:0.5) ..  (QQ.north east);
\draw[<->,color=Orange,line width=1.5] (SS.south) to ([yshift=.2em]NN.north);
\draw[<->,color=Orange,line width=1.5] ([yshift=+.2em]TT.south) to ([yshift=0em]MM.north);
\draw[<->,color=Orange,line width=1.5] (UU.south) to ([yshift=.2em]LL.north);
}
\end{table}

\paragraph{Application to the Intriligator-Seiberg duality:}
Our result thus far is equally applicable in the case of \Nequals1 supersymmetric QCD, 
since they are connected to the non-supersymmetric QCD by a continuous deformation 
preserving all the symmetries we care about.
Now, let us recall that Intriligator and Seiberg found in \cite{Intriligator:1995id} a duality exchanging $\so(N_c)$ and $\so(N_f-N_c+4)$,
which in our notation sends $n_c$ to $n_c'=n_f-n_c+2$, keeping $n_f$ fixed.

Following a crucial set of observations in \cite{Strassler:1997fe} that spinors in the original theory are mapped to magnetic monopoles in the dual theory,
the Intriligator-Seiberg duality of \Nequals1 $\so$ theories was refined in \cite{Aharony:2013hda},
to account for the global form of the gauge group and the discrete theta angle.
It was concluded there that $\Spin$ is exchanged with $\SO_-$ while $\SO_+$ maps to itself.
This mapping was given a further confirmation by using supersymmetric localization on $S^3/\bZ_n \times S^1$ in \cite{Razamat:2013opa}.
Our analysis allows us to check this duality by comparing how the 0-form symmetry and the 1-form symmetry are combined in the dual pairs.
We superimposed the action of the duality on our main Table~\ref{table:main} for the massless case
and Table~\ref{table:sub} for the massive case.
It is satisfying to see that the duality action correctly preserves the behaviors `none', `anomaly' and `extension'.
In the last couple of years, the study of higher symmetries and their anomalies of supersymmetric theories has seen some activity,\footnote{%
See e.g.~\cite{Cordova:2020tij,DelZotto:2020sop,Gukov:2020btk,Apruzzi:2021vcu,Bhardwaj:2021wif} where  2-groups of supersymmetric theories were studied.}
but mostly from the point of view of string theory or M-theory.
The authors hope that this paper paves a way toward a more field-theoretical analysis of these matters.

\medskip

\paragraph{Organization of the paper:}
The rest of the paper is organized as follows.
In Sec.~\ref{sec:2-group},
we determine exactly when the electric\,/\,magnetic\,/\,dyonic $\bZ_2$ 1-form symmetries
and the $\SU(2n_f)/\bZ_2$ flavor 0-form symmetry form a nontrivial 2-group,
by examining the charges of line operators in each theory.
In Sec.~\ref{sec:sl2z},
we exploit the $\SL(2,\bZ_2)$ actions on theories with $\bZ_2$ 1-form symmetries, including our $\so(2n_f)$ QCDs.
This will allow us to determine the 't Hooft anomalies they possess.
Combining the results with those obtained in Sec.~\ref{sec:2-group},
one can completely determine the structures of symmetries and anomalies of $\so(2n_f)$ QCDs,
and can further confirm that they are indeed compatible with the Intriligator-Seiberg duality.
Although the result itself is satisfactory, the analysis leading to it is somewhat ad-hoc, 
so in Sec.~\ref{sec:fermion}, we partially complement it with a more direct computation of fermion anomalies.

In Appendix~\ref{sec:remark}, we discuss how we can understand the 2-group structure in general by studying line operators and line-changing point operators,
and find a relation to the crossed module extensions classifying $H^3$.
Finally, we have the two appendices providing technical details of the mathematical facts used in the main part;
in Appendix~\ref{sec:bordism}, we compute relevant bordism groups capturing the anomalies of spin QFTs associated with various symmetries;
and in Appendix~\ref{sec:nonclosedP}, we describe some subtleties concerning the Pontrjagin square.

Before proceeding, we list the obstruction classes which will be frequently encountered in this paper.
In general, given a group $G$, a subgroup $\bZ_n$ in the center of $G$,
and a $G/\bZ_n$ bundle on a manifold $X$, 
there is a obstruction class in $H^2(X; \bZ_n)$ controlling whether this bundle lifts to a $G$ bundle.
For $G=\Spin(N_c)$ and $G/\bZ_2=\SO(N_c)$ this is the familiar second Stiefel-Whitney class $w_2$.
The classes we use are listed  in Table~\ref{table:w2}.

\begin{table}[h]
\[
\begin{array}{c| rr| c|c}
\text{name} & \multicolumn{1}{c}{G/\bZ_n} & \multicolumn{1}{c|}{G} & \bZ_n & \text{comments}\\
\hline 
w_2 & \SO(2n_c)\phantom{/\bZ_2} & \Spin(2n_c) & \bZ_2\\
v_2 & \SO(2n_c)/\bZ_2 & \SO(2n_c) &\bZ_2\\
x_2 & \SO(2n_c)/\bZ_2 & \Spin(2n_c) &\bZ_4 & n_c:\text{odd}\\
\hline
a_2 & \SU(2n_f)/\bZ_2 & \SU(2n_f) &\bZ_2\\
a_2 & \USp(2n_f)/\bZ_2 & \USp(2n_f) &\bZ_2\\
a_2 & \U(n_f)/\bZ_2 & \U(n_f) &\bZ_2\\
\end{array}
\]
\caption{The names we use for the obstruction classes $\in H^2(X,\bZ_n)$ controlling whether a $G/\bZ_n$ bundle
on $X$ lifts to a $G$ bundle.  \label{table:w2}}
\end{table}

\section{2-group structure}
\label{sec:2-group}
Let us first study whether the $\bZ_2$ 1-form symmetry and the $\SU(2n_f)/\bZ_2$ 0-form flavor symmetry form a nontrivial 2-group or not. 
This can be found rather physically by studying the line operators. 

\subsection{$\Spin$}
We start by discussing the $\Spin(2n_c)$ gauge theory with $2n_f$ fermions in the vector representation. 
The results presented in this subsection was originally found in \cite[Sec.~4.4]{Hsin:2020nts}.

First, recall that the center of $\Spin(2n_c)$ is $\bZ_2\times \bZ_2$ or $\bZ_4$ depending on whether $n_c$ is even or odd.
This corresponds to the fact that the tensor square of a spinor representation contains the identity representation when $n_c$ is even while it contains the vector representation when $n_c$ is odd.

Now, consider the Wilson line $W$ in the spinor representation.
When $n_c$ is even, $W^2$ contains the identity representation, and therefore we simply have a $\bZ_2$ 1-form symmetry independent of the flavor symmetry, and there is nothing more to see here.

When $n_c$ is odd, $W^2$ contains the vector representation.
This can be screened by the dynamical fermion, which however carries the fundamental representation of $\SU(2n_f)$ flavor symmetry, 
and in particular transforms nontrivially under $-1\in \SU(2n_f)$.
In other words, the flavor Wilson line in the fundamental representation of $\SU(2n_f)$ can now be considered as the square  of the gauge Wilson line in the spinor representation of $\Spin(2n_c)$.
This means that we have the following extension of groups \begin{equation}
	0
	\quad
	\longrightarrow \underbrace{\bZ_2}_{\substack{
		\text{group of}\\
		\text{charges under}\\
		\text{$\{\pm1\}\in \SU(2n_f)$}
	}}
	\longrightarrow \quad \bZ_4 
	\quad
	\longrightarrow \underbrace{\bZ_2}_{\substack{
		\text{group of}\\
		\text{gauge Wilson lines}\\
		\text{up to screening}
	}}
	\longrightarrow \quad 0.
\label{charge-extension}
\end{equation}
As the groups of charges of $\SU(2n_f)$ 0-form symmetry and $\bZ_2$ 1-form symmetry are combined nontrivially, 
the symmetry groups themselves are also combined nontrivially.
Let us see this point by considering their background fields.
(We will discuss another general method  to relate this extension to 2-groups in Appendix~\ref{sec:remark}.)

The fermion fields are simultaneously in the vector representation of the gauge $\so(2n_c)$ and the fundamental representation of the flavor $\su(2n_f)$,
and therefore are in a representation of $G=\tfrac{\SO(2n_c)\times \SU(2n_f)}{\bZ_2}$.
Given a $G$ bundle on a manifold $X$,
there is an $\SO(2n_c)/\bZ_2$ bundle and an $\SU(2n_f)/\bZ_2$ bundle associated with it.
Let us denote by $v_2,a_2\in H^2(X;\bZ_2)$ the obstruction classes controlling whether they lift to an $\SO(2n_c)$ bundle and an $\SU(2n_f)$ bundle respectively. 
Then we have $v_2=a_2$ for a $G$ bundle.
The flavor Wilson line in the fundamental representation is charged under $-1\in \SU(2n_f)$ in the center,
and $a_2$ can be considered as the background field for this $\bZ_2$ 1-form symmetry.

Now, without the flavor background,  the background $E\in H^2(X;\bZ_2)$ for the electric $\bZ_2$ 1-form symmetry of the $\Spin(2n_c)$ theory sets the Stiefel-Whitney class $w_2\in H^2(X;\bZ_2)$ of the $\SO(2n_c)$ gauge bundle to be $E=w_2$, which controls whether it lifts to a $\Spin(2n_c)$ bundle.
When the flavor background $a_2$ is nontrivial,
the obstruction class $v_2$ controlling the lift from an $\SO(2n_c)/\bZ_2$ bundle to an $\SO(2n_c)$ bundle is nontrivial.
In this situation, when $n_c$ is odd, $w_2$ can no longer be defined as a closed cochain; rather it satisfies $\delta w_2 = \beta v_2$, where $\beta$ is the Bockstein operation,
since\footnote{%
Indeed, let $\overline{v}_2$ the $\bZ_4$-lift of the cochain $v_2$, where the value $\{0,1\}$ are lifted to $\{0,1\}\subset \{0,1,2,3\}$. 
$\beta v_2$ is by definition $\tfrac12 \delta \overline{v}_2$, as we explained in footnote \ref{foot:bock}.
The $\bZ_2$-reduction of the cochain $x_2$ is $v_2$, and $x_2-\overline{v}_2$ is divisible by 2,
so we can identify $w_2 = \tfrac12 (x_2-\overline{v}_2)$.
As $\tfrac12\delta x_2$ is zero as a $\bZ_2$-valued cochain, we find $\delta w_2=\beta v_2$, as desired.
} together they specify the obstruction class $x_2\in H^2(X;\bZ_4)$ 
controlling the lift from an $\SO(2n_c)/\bZ_2=\Spin(2n_c)/\bZ_4$ bundle to a $\Spin(2n_c)$ bundle.
As $E=w_2$ and $v_2=a_2$, we conclude that the background field satisfies \begin{equation}
\delta E=\beta a_2.
\label{ba}
\end{equation}

In general, a 2-group $H$ combining a 1-form symmetry $A$ and a 0-form symmetry $G$,
which fits in the exact sequence \begin{equation}
	0
	\longrightarrow A[1]
	\longrightarrow H
	\longrightarrow G
	\longrightarrow 0
\end{equation} with the extension class $\alpha\in H^3(BG;A)$,
is defined as a symmetry whose background field is given by a pair of a degree-2 cochain $E\in C^2(X,A)$ and a background $G$ field $g:X\to BG$ satisfying $\delta E = g^*(\alpha)$.
Here $A[1]$ means the Abelian group $A$ regarded as a 1-form symmetry,
and we drop the pull-back symbol $g^*$ when its presence is clear from the context.
In our case, we see that the $\bZ_2$ 1-form symmetry and the $\SU(2n_f)/\bZ_2$ 0-form flavor symmetry form the 2-group $H$ fitting in the sequence \begin{equation}
	0
	\longrightarrow \bZ_2[1]
	\longrightarrow H
	\longrightarrow \SU(2n_f)/\bZ_2
	\longrightarrow 0
\label{group-extension}
\end{equation} with the extension class being $\beta a_2 \in H^3(B(\SU(2n_f)/\bZ_2);\bZ_2)$.\footnote{%
When the 0-form symmetry part is finite, the extension class can be visualized in terms of 
the codimension-2 operator implementing the 1-form symmetry,
emerging from the triple-intersections of three codimension-1 operator implementing 0-form symmetry,
see \cite{Benini:2018reh}.
}

Note that having the extension of groups of charges of line operators as in \eqref{charge-extension}
is equivalent to having a nontrivial 2-group extension \eqref{group-extension} 
whose background field satisfies \eqref{ba}.
The situation can be summarized as the following commuting diagram:
\begin{equation}
\begin{tikzcd}
& &0\arrow[d]&\arrow[d]0& \\
 &0\arrow[r]\arrow[d]&\SU(2n_f)\arrow[r]\arrow[d]&\SU(2n_f)\arrow[d]\arrow[r]& 0 \\
 0 \arrow[r]& \bZ_2[1] \arrow [r]\arrow[d] & H \arrow[r] \arrow[d] & \SU(2n_f)/\mathbb{Z}_2 \arrow[r]\arrow[d,"w_2"] & 0\\
 0 \arrow[r]& \bZ_2[1] \arrow [r]\arrow[d] & A_H[1] \arrow[r]\arrow[d] & \bZ_2[1] \arrow[r]\arrow[d] & 0\\
 &0&0&0&
\end{tikzcd}.
\end{equation}
Here, the sequences of the form $0\to G\to G'\to G''\to 0$ in the columns and the rows 
are to be interpreted as having fibration sequences $BG\to BG'\to BG''$ among the respective classifying spaces.\footnote{%
In particular, the maps $\SU(2n_f)\to \SU(2n_f)/\bZ_2$ and $\SU(2n_f)\to H$ are not injective in the usual sense.
}
We note that  the map $w_2:\SU(2n_f)/\bZ_2 \to \bZ_2[1]$ extracts the information of the obstruction class $a_2 \in H^2(B(\SU(2n_f)/\bZ_2);\bZ_2)$.
We also note that the 2-group $H$ is uniquely determined by $A_H[1]$:
if $A_H[1] = \mathbb{Z}_4[1]$, the extension is nontrivial, while $A_H[1] = (\bZ_2\times \bZ_2)[1]$, it is trivial.
Therefore, to determine the 2-group extension, 
we can simply study the group of charges $A_H$ of line operators,
which we will carry out for $\SO_\pm$ gauge theories next.

\subsection{$\SO_\pm$}

We would like to study how the magnetic\,/\,dyonic $\bZ_2$ 1-form symmetry of the $\SO(2n_c)_\pm$ gauge theory is combined with the $\SU(2n_f)$ flavor symmetry.
We first discuss the case $\SO_+$ in detail; the minor changes needed to take $\SO_-$ into account would be described later.

In accord with the discussions in the previous subsection, 
we consider what happens when we take two copies of the 't Hooft line operator $H$ and fuse them.
At the very naive level, $H^2$ can be screened by dynamical monopoles,
but dynamical monopoles can receive flavor\,/\,gauge center charges from the fermion zero modes.

\paragraph{Making deformations:}
To study these issues, it is useful to deform the theory and make it simpler
by performing the following steps:
\begin{itemize}
\item Reduce the flavor symmetry from $\SU(2n_f)$ to $\USp(2n_f)$. 
The fundamental representation still transforms nontrivially under $-1\in \USp(2n_f)$, which is enough for our purposes.
\item Add an adjoint scalar $\Phi_{[ab]}$ and the interaction $\psi^{ai}_\alpha \psi^{bj}_\beta J_{ij} \Phi_{ab}\epsilon^{\alpha\beta} +c.c.$.
Here $a,b$ and $i,j$ are vector indices of $\SO(2n_c)$ and $\USp(2n_f)$,
$\alpha,\beta$ are the spinor indices,
and  $J_{[ij]}$ is the constant invariant matrix for the $\USp(2n_f)$.
\item Give a generic vacuum expectation value (vev) to $\Phi_{ab}$ and break $\SO(2n_c)$ to $\SO(2)^{n_c}$.
\end{itemize}

The 't Hooft lines in the resulting $\SO(2)^{n_c}$ theory can be labeled by their magnetic charges $(m_1,\ldots,m_{n_c})\in \bZ^{n_c}$.
The dynamical monopoles have the charges in the `adjoint class', which are
in the root lattice $\Lambda$ of $\SO(2n_c)$. 
Then, the group of the magnetic charges of 't Hooft lines up to screening by the dynamical monopoles is \begin{equation}
\bZ^{n_c}/\Lambda = \bZ_2,
\end{equation}
which agrees with the 1-form symmetry before the deformation.
We now would like to study how this $\bZ_2$ is combined with the flavor\,/\,gauge center $\bZ_2$ charge.

\paragraph{Reduction to  the $\so(4)$ case:}
For this purpose we need to know slightly more details of the dynamical monopoles.
The dynamical monopoles associated with the breaking of a gauge group to its Cartan were analyzed in many places, 
e.g.~in \cite{Weinberg:1979zt}.
There, the following was shown.
Let $\phi$ be the scalar vev in the real Cartan subalgebra, $\phi\in \mathfrak{h}\subset \mathfrak{g}$.
This determines the simple roots $\alpha$.
Then you can embed the standard spherically-symmetric 't Hooft-Polyakov monopole
using the $\mathfrak{su}(2)$ subalgebra associated with $\alpha$,
and have a monopole solution without additional bosonic moduli. 

Let us say we chose the standard $\phi$ such that the simple roots are 
\begin{equation}
(1,-1,\ldots,0),\ 
(0,1,-1,\ldots,0),\ldots,
(0,\ldots,1,-1),\ 
(0,\ldots,1,+1)\in \bZ^{n_c},
\end{equation}
which we call simple dynamical monopoles.
Now, consider the group $\bZ^{n_c}\times \bZ_2$ which combines the magnetic charges in $\bZ^{n_c}$ and the flavor\,/\,gauge center charge $q \in \bZ_2$.
What we are after is the quotient of $\bZ^{n_c}\times \bZ_2$ 
by the subgroup generated by the charges of simple dynamical monopoles,
which we denote respectively by
\begin{equation}
(1,-1,\ldots,0;q_1),\ 
(0,1,-1,\ldots,0;q_2),\ldots,
(0,\ldots,1,-1;q_{n_c-1}),\ 
(0,\ldots,1,+1;q_{n_c}).
\end{equation}
To determine this quotient,
we do not have to determine the all $q_i$'s; 
we simply use the first $n_c-2$ vectors to relate any charge vector $
(m_1,\ldots, m_{n_c-2}, m_{n_c-1},m_{n_c}; q)
$
to a vector of the form $
(0,\ldots,0, m, m' ; q').
$
Then, only $q_{n_c-1}$ and $q_{n_c}$ need to be determined.
This reduces the study to the case of $n_c=2$ and $\so(2n_c) = \so(4) \simeq \su(2)_1 \times \su(2)_2$,
where the monopoles associated with the simple roots are just 't Hooft-Polyakov monopoles associated with the two factors of $\su(2)$'s.

\paragraph{Analysis of the $\so(4)$ case:}

The vev of the adjoint scalar in this basis can be written as $(a_1,a_2)$, which we assume to be $a_1>a_2>0$.
Here, the fermion is in the vector representation of $\so(4)$.
Under the monopole in $\su(2)_1$, it is a doublet coupled to an adjoint vev of size $a_1$ with bare mass $a_2$,
and similarly for the monopole in $\su(2)_2$.

Now, the explicit analysis in \cite[Sec.~IV]{Callias:1977kg} concerning the number of zero modes in the \mbox{'t Hooft}-Polyakov monopole says that 
a doublet fermion coupled to an adjoint vev of size $a$ with bare mass $\mu$ has
a zero mode if $|a|>|\mu|$ and 
has no zero modes if $|a|<|\mu|$.
With our assumption $a_1>a_2>0$, this means that the monopole in $\su(2)_1$ has a zero mode, while the monopole in $\su(2)_2$ does not.
In our original basis, this means that the monopole with $(0,\ldots,1,-1;q_{n_c-1})$ does not produce any zero modes and $q_{n_c-1}=0$,
while the monopole with $(0,\ldots,1,+1;q_{n_c})$  has two zero modes per flavor.
The 1-form symmetry group is obtained by dividing $\bZ^2\times \bZ_2$ by the subgroup generated by $(1,-1;0)$ and $(1,+1;q_{n_c})$.
This is $\bZ_2\times \bZ_2$ or $\bZ_4$ depending on whether $q_{n_c}$ is $0$ or $1$.

Let us determine $q_{n_c}$, the center charge of the monopole in $\su(2)_1$.
We saw that there are two zero modes per flavor; 
this means that there are fermionic zero modes transforming in \begin{equation}
R_{2n_f} \otimes V_{2},
\end{equation} where $R_{2n_f}$ is the fundamental representation of $\usp(2n_f)$,
while $V_2$ is the doublet of $\su(2)_2$,\footnote{
	It is actually broken to $\u(1)$, but keeping $\su(2)_2$ representation is useful in organizing the answer.
}
and we need to impose the reality condition using the pseudo-reality of both factors,
so that there are $4n_f$ Majorana fermion in total.

To determine the flavor\,/\,gauge center charge $q_{n_c}$ of the monopole,
it suffices to consider the case $n_f=1$;
the general case is given simply by multiplying it by $n_f$.
When $n_f=1$, there are $4$ Majorana fermions.
Quantizing them, we find the monopoles in \begin{equation}
(R_2\otimes \mathbf{1}) \oplus (\mathbf{1} \otimes V_2).
\end{equation}
It has the `vector' charge under $\usp(2)\simeq\su(2)$ flavor symmetry or is a doublet under $\su(2)_2$,
which corresponds to the `vector' charge under $\so(4)$ gauge symmetry.
In either case, they have the flavor\,/\,gauge center charge $1 \in \{0,1\}=\bZ_2$.
Therefore we conclude the flavor\,/\,gauge center charge $q_{n_c}$ is simply given by $n_f$ mod 2.

\paragraph{Summary:}
Combining the intermediate steps we described above, we conclude the following: 
for the $\SO(2n_c)_+$ gauge theory,
the group $\bZ_2$ of magnetic charges of 't Hooft lines is extended by the flavor\,/\,gauge center symmetry $\bZ_2$ to become $\bZ_4$ when $n_f$ is odd,
while they remain separate when $n_f$ is even.

The analysis of the $\SO(2n_c)_-$ gauge theory is largely the same;
the only difference is that the discrete theta angle gives an additional gauge center charge\footnote{%
To see this, note that the original interaction $2\pi i\int\tfrac14 \fP(w_2)$ induces the interaction $2\pi i \int \tfrac14 (\sum_{i=1}^{n_c} c_1^{(i)} )^2$ in the $\SO(2)^{n_c}$ theory.
This gives the electric charge $(1,1,\ldots,1)$ to the monopole with the magnetic charge $(0,0,\ldots,1,+1)$.
Under $-1\in SO(2n_c)$ such a state transforms by $(-1)^{n_c}$.
} to the simple dynamical monopole with the magnetic charge $(0,0,\ldots,1,+1)$, so that $q_{n_c}=n_f+n_c$ mod 2.
Therefore, we conclude the following:
for the $\SO(2n_c)_-$ gauge theory,
the group $\bZ_2$ of magnetic charges of 't Hooft lines is extended by the flavor\,/\,gauge center symmetry $\bZ_2$ to become $\bZ_4$ when $n_f+n_c$ is odd,
while they remain separate when $n_f+n_c$ is even.

The result of the analysis is summarized in Table~\ref{table:2group}.
There, `product' means that the $\bZ_2$ 1-form symmetry and the $\SU(2n_f)/\bZ_2$ flavor symmetry are kept separate and form a direct product,
while `extension' means that they form a nontrivial 2-group.
We remark that the nontrivial 2-group is always given by the extension \eqref{group-extension}  whose background fields satisfy \eqref{ba}.

\begin{table}[h]
\centering
\renewcommand{\arraystretch}{1.2}
\begin{tabular}{r@{\,}l|ccc}
$(n_c,$&$n_f)$ & $\Spin$ & $\SO_+$ & $\SO_-$\\
\hline
(even,&even) & \textcolor{lesslightgray}{product} & \textcolor{lesslightgray}{product} & \textcolor{lesslightgray}{product} \\
(odd,&even) & extension & \textcolor{lesslightgray}{product} & extension \\
(even,&odd) & \textcolor{lesslightgray}{product} & extension & extension \\
(odd,&odd) & extension & extension & \textcolor{lesslightgray}{product}
\end{tabular}
\caption{How the $\bZ_2$ 1-form symmetry and the $\SU(N_f)/\bZ_2$ 0-form flavor symmetry are combined
in $\so(2n_c)$ QCD.
The label `product' means that they form a direct product,
while the label `extension' means that they form a nontrivial 2-group.
\label{table:2group}}
\end{table}

\section{$\SL(2,\bZ_2)$ action and the anomalies}
\label{sec:sl2z}
In the last section we determined the 2-group structure of the $\so(2n_c)$ gauge theories with $2n_f$ flavors, by studying the group of the charges of line operators. 
Here we determine the anomalies of these symmetries, 
utilizing the $\SL(2,\bZ_2)$ action on the set of QFTs with $\bZ_2$ 1-form symmetry.

\subsection{$\SL(2,\bZ_2)$ action and $\so$ gauge theories}

Let us say that we are given a four-dimensional spin QFT $Q$ with $\bZ_2$ 1-form symmetry.
We denote its partition function on a manifold $X$ by $Z_Q[E]$, 
where we suppress the dependence on $X$ in the notation, and $E\in H^2(X;\bZ_2)$ is the background field for the $\bZ_2$ 1-form symmetry.
We then define $SQ$ and $TQ$ to be QFTs with partition functions given by the formula \begin{equation}
Z_{SQ}[B] \propto \sum_{E} (-1)^{\int_X B\cup E} Z_Q[E],\qquad
Z_{TQ}[E]= (-1)^{\int_X \tfrac12 \fP(E)} Z_Q[E],
\end{equation}
where $\fP: H^2(-;\bZ_2) \to H^4(-;\bZ_4)$ is a cohomology operation called the Pontrjagin square.
We can show that $S^2=T^2=1$ and $(ST)^3=1$, meaning that they generate $\SL(2,\bZ_2)$.
This operation was introduced in  \cite{Gaiotto:2014kfa} as an analogue of the $\SL(2,\bZ)$ action on 3d QFTs with $\U(1)$ symmetry of  \cite{Witten:2003ya} and then further studied in \cite{Bhardwaj:2020ymp}.

Importantly, $\Spin(2n_c)$ and $\SO(2n_c)_\pm$  gauge theories with $2n_f$ flavors with the same $(n_c,n_f)$ form a single orbit under this $\SL(2,\bZ_2)$ action.
More precisely, we need to make a distinction between $\Spin(2n_c)$ and $T(\Spin(2n_c))$, and similarly between $\SO(2n_c)_\pm$ and $T(\SO(2n_c)_\pm)$ respectively,
where the theories with $T$ prepended are different from the original ones only by its discrete theta coupling to the background.
Then we have the following chain of actions: \begin{equation}
\begin{tikzpicture}[baseline=(BASE.base)]
	\node (BASE) at (0,0) {$
		T(\Spin) \stackrel{T}{\longleftrightarrow} 
		\Spin \stackrel{S}{\longleftrightarrow} 
		\SO_+ \stackrel{T}{\longleftrightarrow} 
		T(\SO_+) \stackrel{S}{\longleftrightarrow} 
		T(\SO_-) \stackrel{T}{\longleftrightarrow} 
		\SO_-.
	$};
	\draw[->] (-5,-1) -- (-5,-0.5);
	\draw[->] (5.4,-1) -- (5.4,-0.5);
	\draw (-5,-1) -- node[above] {\scriptsize $S$} (5.4,-1);
\end{tikzpicture}
\label{sl2z-on-so}
\end{equation}

\subsection{$\SL(2,\bZ_2)$ actions with extra background}

Let us now study what happens if we perform this $\SL(2,\bZ_2)$ action when the $\bZ_2$ 1-form symmetry in question is part of a larger symmetry group.
So far we have been considering the effect of $\SU(2n_f)/\bZ_2$ 0-form flavor symmetry,
but the discussions in the last section show that, at a formal level, only the background field $a_2 \in H^2(X;\bZ_2)$ matters, which controls the lift from an $\SU(2n_f)/\bZ_2$ bundle to an $\SU(2n_f)$ bundle.
Let us regard $a_2$ as the background field for a flavor $\bZ_2$ 1-form symmetry.

Then, it is combined with the original $\bZ_2$ 1-form symmetry into either $\bZ_2\times \bZ_2$ or $\bZ_4$,
and we perform the $\SL(2,\bZ_2)$ action  by picking a $\bZ_2$ subgroup.
The symmetry and the anomaly of the resulting theory can be determined by a formal argument independent of the dynamics of the theory,
once those of the original theory 
and the action of the anomaly-free subgroup to be gauged are given, as discussed in \cite{Tachikawa:2017gyf}.

Let us work at the level of anomalies described by cohomology, since we do not need to deal with more general anomalies described by bordism.
We consider a $d$-dimensional QFT with a symmetry group $G$ with an anomaly specified by a cochain $\alpha\in C^{d+1}(BG;\U(1))$. 
We pick a subgroup $H\subset G$ such that $\alpha$ trivializes in it, so that one can find its trivialization $\mu \in C^d(BH;\U(1))$ satisfying $\alpha|_H = \delta \mu$. 
We then gauge $H$, using $\mu$ as the action.

What determines the symmetry and the anomaly of the gauged theory is the data $(\mu,\alpha)$.
Clearly, given $\nu\in C^d(BG;\U(1))$, the pair $(\mu,\alpha)$ and the pair $(\mu-\nu|_H,\alpha-\delta\nu)$ should give the same result, since we merely added the counterterm $\nu$ to the action.
This allows us to always choose the pair of the form $(0,\alpha')$ equivalent to a given $(\mu,\alpha)$, 
by taking $\nu$ to be an arbitrary lift of $\mu$ from $H$ to $G$.
This is convenient in discussing the $\SL(2,\bZ_2)$ action,
since our $S$ operation is defined in the convention that $\mu=0$.

At this stage, the residual identifications $(0,\alpha')\sim (0,\alpha'')$ are of the form 
$\alpha''=\alpha'+\delta \nu$, where $\nu\in C^{d}(BG;\U(1))$ is required to satisfy $\nu|_H=0$.
Their equivalence classes form the relative cohomology group $H^{d+1}(BG,BH;\U(1))$.\footnote{
It might be interesting to study anomalies taking values in the relative cohomology (or bordism) groups.
}

\paragraph{The four choices:}
Now, what are the possible choices of $(\mu,\alpha)\sim(0,\alpha')$ we need to discuss?
Let us first consider $\bZ_2\times \bZ_2$ 1-form symmetry.
As detailed in the Appendix~\ref{sec:bordism}, the only possible anomaly for 4d spin QFTs with this symmetry is \begin{equation}
\alpha=
\frac12  B\beta E , \label{anom}
\end{equation} where $B,E\in H^2(Y;\bZ_2)$ are the background fields on the bulk 5d spin manifold $Y$,
and we use $\bQ/\bZ$-valued cochains to describe the anomaly.
Its restriction to $\bZ_2$ 1-form symmetry subgroup is trivial i.e.~$\alpha|_{H=\bZ_2} = 0$,
and thus the possible choice of $\mu$ is simply the discrete theta angle \begin{equation}
\mu =
 \frac14 \fP(E), \label{P}
\end{equation}
where $\fP$ is the Pontrjagin square.
This $\mu$ can be lifted from the $\bZ_2$ subgroup to the entire $\bZ_2\times \bZ_2$ group as a closed cochain, 
and therefore does not affect the gauging process.
Therefore, we only have to consider pairs $(0,0)$ and $(0,\alpha)$.

Next, we consider $\bZ_4$ 1-form symmetry.
In the Appendix~\ref{sec:bordism}, we show that there is no anomaly for $\bZ_4$ 1-form symmetry.
Therefore we can pick $\alpha=0$. Then the only possible choice of $\mu$ for the $\bZ_2$ 1-form subgroup is again the discrete theta angle \eqref{P}.
One difference here is that the discrete theta angle \eqref{P} cannot be lifted as a closed cochain to the entire $\bZ_4$ 1-form subgroup.
As discussed in the Appendix~\ref{sec:nonclosedP}, with $\delta E=\beta a_2$
where $a_2\in H^2(X;\bZ_4/\bZ_2)$,
one finds \begin{equation}
\alpha':=\delta \mu = 
\frac12 a_2 \beta_2 \tilde a_2, \label{anom'}
\end{equation}
where $\beta_2$ is the higher Bockstein operation associated with
the short exact sequence
\begin{equation}
	0
	\longrightarrow \bZ_2
	\longrightarrow \bZ_8
	\longrightarrow \bZ_4
	\longrightarrow 0,
\end{equation}
and $\tilde a_2$ is the lift of $a_2$ to a $\bZ_4$-valued cochain;
see also footnote~\ref{foot:bock}.
We conclude that the pairs  we need to consider for the $\bZ_4$ 1-form symmetry are $(0,0)$ and $(\mu,0)\sim (0,\alpha')$. 

\def\Textended{extended$_T$}
Summarizing,  we need to consider the following four choices, namely:
\begin{itemize}
\item For $\bZ_2\times \bZ_2$, the pairs $(0,0)$ and $(0,\alpha)$, which we call `none' and `anomaly'
\item For $\bZ_4$, the pairs $(0,0)$ and $(\mu,0)\sim (0,\alpha')$, which we call `extended' and `\Textended'.
\end{itemize}

\paragraph{$\SL(2,\bZ_2)$ action on the four choices:}
Let us now determine how the $\SL(2,\bZ_2)$ action affects these data.
The case `none' is very easy.
The additional $\bZ_2$ factor plays no role, and we find the chain of actions given by \begin{equation}
\begin{tikzpicture}[baseline=(BASE.base)]
	\node (BASE) at (0,0) {$
		\text{none} \stackrel{T}{\longleftrightarrow} 
		\text{none} \stackrel{S}{\longleftrightarrow} 
		\text{none} \stackrel{T}{\longleftrightarrow} 
		\text{none} \stackrel{S}{\longleftrightarrow} 
		\text{none} \stackrel{T}{\longleftrightarrow} 
		\text{none}.
	$};
	\draw[->] (-4.6,-1) -- (-4.6,-0.5);
	\draw[->] (4.5,-1) -- (4.5,-0.5);
	\draw (-4.6,-1) -- node[above] {\scriptsize $S$} (4.5,-1);
\end{tikzpicture}
\label{trivial-chain}
\end{equation}
In the rest of this subsection, we will establish the chain of actions \begin{equation}
\begin{tikzpicture}[baseline=(BASE.base)]
	\node (BASE) at (0,0) {$
		\text{\Textended} \stackrel{T}{\longleftrightarrow} 
		\text{extended} \stackrel{S}{\longleftrightarrow} 
		\text{anomaly} \stackrel{T}{\longleftrightarrow} 
		\text{anomaly} \stackrel{S}{\longleftrightarrow} 
		\text{extended}\stackrel{T}{\longleftrightarrow} 
		\text{\Textended}.
	$};
	\draw[->] (-6.5,-1) -- (-6.5,-0.5);
	\draw[->] (6.3,-1) -- (6.3,-0.5);
	\draw (-6.5,-1) -- node[above] {\scriptsize $S$} (6.3,-1);
\end{tikzpicture}
\label{nontrivial-chain}
\end{equation}

We already explained above that $T$ (i.e.~adding the discrete theta angle \eqref{P}) leaves `anomaly' unchanged,
while it exchanges `extended' and `\Textended'.
To establish the chain above, we then need to show that 
$S$ exchanges `extended' and `anomaly'
while leaves `\Textended' unchanged.

That $S$ exchanges `extended' and `anomaly'
was in fact already reviewed in the Introduction, around \eqref{coupling} and \eqref{mixed},
where we started from `extended', 
gauged the $\bZ_2$ subgroup of $\bZ_4$, and found the `anomaly',
as first demonstrated in \cite{Tachikawa:2017gyf}.

That $S$ leaves `\Textended' unchanged was established in \cite{Hsin:2020nts}.
We will provide a slightly different explanation than the one given there.
Recalling that `\Textended' can be obtained by performing the $T$ transformation on `extended',
its $S$ transformation then involves the coupling
 \begin{equation}
\exp\left[2\pi i\int_X \left( \frac12 B E + \frac14 \fP(E)\right)\right],
\label{boo}
\end{equation}
where $E$ is the variable to be gauged and $B$ is the newly introduced background field.
As $\bZ_2$ to be gauged is the $\bZ_2$ subgroup of $\bZ_4$ 1-form symmetry,
$E$ is not necessarily closed, but rather satisfies the relation \begin{equation}
\delta E = \beta a_2,
\end{equation}
where $a_2$ is the background field for the quotient $\bZ_4/\bZ_2$ 1-form symmetry.
Then the second term in \eqref{boo} is not closed, 
and to even talk about the first term in \eqref{boo}, one first needs to extend the definition of the Pontrjagin square $\fP$ to non-closed cochains,
as we discuss in Appednix~\ref{sec:nonclosedP}.

To make  the coupling \eqref{boo}  well-defined, we consider adding a counterterm $\tfrac14\fP(B)$ depending solely on the newly introduced field $B$ to \eqref{boo},
i.e.~we perform a further $T$ transformation. 
The total coupling is now \begin{equation}
\exp\left[2\pi i\int_X \left( \frac12 B E + \frac14 \fP(E) + \frac14 \fP(B) \right)\right]
= 
\exp \left[2\pi i \int_X \frac14 \fP(E+B)\right] .
\label{E+B}
\end{equation}
This theory is perfectly well-defined and has no anomaly,
if the newly-introduced background field $B$  also satisfies \begin{equation}
\delta B=\beta a_2,
\end{equation}
since $\delta(B+E)=0$.
This means that, starting from `extended' and performing $T$, $S$, and $T$,
we come back to  `extended'.
Therefore, simply performing $S$ for the theory of the type `\Textended',  one  finds `\Textended'.
This establishes the chain of actions shown in \eqref{nontrivial-chain}.

\subsection{Anomalies from $\SL(2,\bZ_2)$ action} 

Let us now combine our result in Table~\ref{table:2group},
which summarizes our knowledge whether the 0-form symmetry and the 1-form symmetry form a nontrivial 2-group,
and the $\SL(2,\bZ_2)$ actions \eqref{trivial-chain} and \eqref{nontrivial-chain} on the four choices  we determined above.
We first need to double each column of Table~\ref{table:2group},
since we need to distinguish $\Spin$ from $T(\Spin)$ and $\SO_\pm$ from $T(\SO_\pm)$.
The entry `product' in Table~\ref{table:2group} corresponds to either `none' or `anomaly',
and the entry `extension' there corresponds to either `extended' or `\Textended'.
We now demand that the $\SL(2,\bZ_2)$ action \eqref{sl2z-on-so} on $\so$ QCD to be compatible with the $\SL(2,\bZ_2)$ action on the labels, \eqref{trivial-chain} and \eqref{nontrivial-chain}.
The only consistent assignment is given in Table~\ref{table:refined}.
As the way we determine the symmetry structures were somewhat indirect,
we confirm the structure of the $\Spin$ case in the next section
in a different means.

\begin{table}
\centering
\[
\begin{array}{r@{\,}l|cccccc}
(n_c,&n_f)  & T(\Spin) & \Spin & \SO_+ & T(\SO_+) & T(\SO_-) & \SO_- \\
\hline
\vphantom{\Bigm|}(\text{even},&\text{even}) & \text{none}\tikzmark{a1}& \color{Purple}\text{none}\tikzmark{a2}& \text{none}\tikzmark{a3}& \tikzmark{a4}\text{none}& \tikzmark{a5}\text{none}& \tikzmark{a6}\text{none}\\
\vphantom{\Bigm|}(\text{odd},&\text{even}) & \text{\Textended}\tikzmark{b1} &\color{Purple}\text{extended}\tikzmark{b2} & \text{anomaly}\tikzmark{b3} & \tikzmark{b4}\text{anomaly} & \tikzmark{b5}\text{extended } & \tikzmark{b6}\text{\Textended} \\
\vphantom{\Bigm|}(\text{even},&\text{odd}) & \text{anomaly}\tikzmark{c1} & \color{Purple}\text{anomaly}\tikzmark{c2} & \text{extended}\tikzmark{c3} & \tikzmark{c4}\text{\Textended}  & \tikzmark{c5}\text{\Textended} & \tikzmark{c6}\text{extended} \\
\vphantom{\Bigm|}(\text{odd},&\text{odd}) & \text{extended}\tikzmark{d1} & \color{Purple}\text{\Textended}\tikzmark{d2} & \text{\Textended}\tikzmark{d3} & \tikzmark{d4}\text{extended}
 &  \tikzmark{d5}\text{anomaly} & \tikzmark{d6}\text{anomaly}   \end{array} 
\]
\tikz[overlay,remember picture]{%
\draw[<->,bend left,color=Orange,line width=1.5] (a1.north) to (a6.north);
\draw[<->,bend left,color=Orange,line width=1.5] (a2.north) to (a5.north);
\draw[<->,bend left,color=Orange,line width=1.5] (a3.north) to (a4.north);
\draw[<->,color=Orange,line width=1.5] (b1.north) .. controls +(10:10em) ..  (b6.north);
\draw[<->,color=Orange,line width=1.5] (b2.north) .. controls +(10:5em) ..  (b5.north);
\draw[<->,color=Orange,line width=1.5] ([yshift=.2em]b3.west) to ([yshift=.2em]b4.east);
\draw[<->,color=Orange,line width=1.5] (c1.south) to (d6.north);
\draw[<->,color=Orange,line width=1.5] (c2.south) to (d5.north);
\draw[<->,color=Orange,line width=1.5] (c3.south) to (d4.north);
\draw[<->,color=Orange,line width=1.5] (c4.south) to (d3.north);
\draw[<->,color=Orange,line width=1.5] (c5.south) to (d2.north);
\draw[<->,color=Orange,line width=1.5] (c6.south) to (d1.north);
}
\caption{The symmetry structure of $\so(2n_c)$ QCD with $2n_f$ flavors,
as deduced from the 2-group structures found in Sec.~\ref{sec:2-group}
and from the $\SL(2,\bZ_2)$ action discussed in this section.
The symmetry structure of the $\Spin$ case, colored in purple, will be checked independently in Sec.~\ref{sec:fermion}.
The action of Intriligator-Seiberg duality is also superimposed using orange arrows.
\label{table:refined}}
\end{table}

We can also use this Table~\ref{table:refined}
to give a further check of the Intriligator-Seiberg duality,
which is known to act as follows, as shown in  \cite[Sec.~6]{Gaiotto:2014kfa}:
\begin{equation}
\renewcommand{\arraystretch}{1.1}
\begin{array}{ccc}
	\Spin(2n_c) & \leftrightarrow & T(\SO_-(2n_f-2n_c+4)),\\
	\SO_+(2n_c) & \leftrightarrow & T(\SO_+(2n_f-2n_c+4)),\\ 
	\SO_-(2n_c) & \leftrightarrow & T(\Spin(2n_f-2n_c+4)).
\end{array}
\end{equation} 
We displayed this action in Table~\ref{table:refined} using orange arrows;
we see that the symmetry structures are indeed preserved across the duality.

\section{Fermion contribution to anomalies}
\label{sec:fermion}
So far, we first determined the 2-group structure in Sec.~\ref{sec:2-group} by studying the charges of line operators,
and then determined the anomalies in Sec.~\ref{sec:sl2z} by matching it to the action of $\SL(2,\bZ_2)$.
Going over  the entries on the column $\Spin$ of Table~\ref{table:refined},
we find that the anomaly is trivial when $(n_c,n_f)$ is (even,\,even) or (odd,\,even),
while it is $\alpha$ given in \eqref{anom} or $\alpha'$ given in \eqref{anom'}
when $(n_c,n_f)$ is (even,\,odd) or (odd,\,odd), respectively.
Since the 1-form symmetry background in the $\Spin$ theory is simply the Stiefel-Whitney class $w_2$ of the $\SO(2n_c)$ gauge bundle, 
these anomalies should simply come from the anomalies of fermions charged under $\tfrac{\SO(2n_c)\times \USp(2n_f)}{\bZ_2}$.
Here we use $\USp(2n_f)$ instead of $\SU(2n_f)$, because under the latter we also have perturbative anomalies, which would complicate the analysis.

For even $n_c$, the anomaly should be given by  \begin{equation}
\alpha=
\frac12 w_2 \beta a_2,
\end{equation}
where $w_2,a_2\in H^2(X;\bZ_2)$ controls the lifts from an $\SO(2n_c)$ bundle to a $\Spin(2n_c)$ bundle
and from a $\USp(2n_f)/\bZ_2$ bundle to a $\USp(2n_f)$ bundle, respectively.
For odd $n_c$, the anomaly cochain should be given by 
 \begin{equation}
\alpha'=
 \frac12 x_2 \beta_2 x_2,
\end{equation} where  $x_2 \in H^2(X;\bZ_4)$ is the class controlling the lift from an $\SO(2n_c)/\bZ_2=\Spin(2n_c)/\bZ_4$ bundle to a $\Spin(2n_c)$ bundle.
We note that, as explained in the previous section, 
 $\alpha'$ is exact as a cocycle on $B(\tfrac{\SO(2n_c)\times \USp(2n_f)}{\bZ_2})$
but defines a nontrivial element in the relative cohomology 
$H^5(B(\tfrac{\SO(2n_c)\times \USp(2n_f)}{\bZ_2}),B\SO(2n_c);\U(1)).$
As such, this cochain still affects the gauging process.

The aim of this last section is to give a check of these anomalies from a different point of view.
We will proceed as follows. 
Starting from the theory where the fermions are charged under $\SO(2n_c)\times \USp(2n_f)$,
we add scalar fields which are adjoint under $\USp(2n_f)$ in the system, and break it down to a subgroup.
We then determine the effective interaction induced by the fermion zero modes.
The next step is to see what happens when the symmetry group is changed from $\SO(2n_c)\times \USp(2n_f)$ to its $\bZ_2$ quotient;
we will see that the effective interaction will have the required anomalies.

Before proceeding, we have two remarks.
First, this method was first used in \cite[Sec.~4]{Witten:1995gf} to understand `a curious minus sign' 
appearing in the topologically-twisted Seiberg-Witten theory, 
which was more recently recognized as determining an anomaly in \cite[Sec.~2.4.3]{Cordova:2018acb}.
It was also used in \cite[Sec.~3.1 and 5.1.2]{Wang:2018qoy} to relate the `new' $\SU(2)$ anomaly 
with the effective interaction in the $\U(1)$ theory.
Second, in this section we can only say that the effective interaction we find is compatible with the anomalies
as found in Sec.~\ref{sec:sl2z},
and will not be able to determine the anomalies completely.
This is mostly due to the fact that the computation of the spin bordism group $\Omega^\text{spin}_d(B\left(\tfrac{\SO(2n_c)\times \USp(2n_f)}{\bZ_2}\right))$ which governs the anomaly is quite hard, 
because even the integral cohomology of the classifying space in question is hard to compute, at least 
to the authors.
Only in a couple of cases we can say more, as we comment along the way.

\subsection{Effective interaction}
We break $\USp(2n_f)$ down to $\U(n_f)$ using a scalar field,
such that the fundamental representation of $\USp(2n_f)$ splits into the fundamental plus the anti-fundamental representation of $\U(n_f)$.
The monopole charge is given by the first Chern class $c_1$ of the low-energy $\U(n_f)$ flavor symmetry.

Take a standard 't Hooft-Polyakov monopole associated with $\U(1)\subset \USp(2)$
and embed it into  $\U(n_f)\subset \USp(2n_f)$.
The fermion zero modes form a vector representation of $\SO(2n_c)$,
whose quantization leads to the spinor representation.
As first discussed in \cite{Thorngren:2014pza} and also used in \cite[Sec.~3.1]{Wang:2018qoy},
this means that there is an effective interaction \begin{equation}
 \frac12 
 w_2(\SO(2n_c)) c_1(\U(n_f)).
 \label{bulk}
\end{equation} 
One way to understand it is as follows.

We started from a system which has $\SO(2n_c)$ symmetry,
but the spinor representation is only a projective representation of this symmetry.
There is an anomaly at the core of the monopole, which needs to flow in from the bulk.
Indeed, taking the spacetime  to be $X=\bR_{\ge 0} \times \bR_t \times S^2$ around the monopole,
and reducing the bulk term \eqref{bulk} on $S^2$ with $\int_{S^2} c_1=1$, 
we have the  effective interaction $\tfrac12\int_Y w_2$
on the half-space $Y=\bR_{\ge 0}\times \bR_t$,
with the monopole living on the boundary. 
Therefore, the degree of freedom on the boundary is in the projective representation characterized by 
$w_2\in H^2(B\SO(2n_c);\bZ_2)$.

\subsection{Anomalies}
We now change the symmetry group from 
$\SO(2n_c)\times \USp(2n_f)$ to $\tfrac{\SO(2n_c)\times \USp(2n_f)}{\bZ_2}$
by taking the $\bZ_2$ quotient.
Note that $\pi_1(\U(n_f)/\bZ_2)=\bZ\times \bZ_2$  or $\bZ$ depending on whether $n_f$ is even or odd.
We denote by $a_2$ the obstruction class to lift a $\U(n_f)/\bZ_2$ bundle to a $\U(n_f)$ bundle.
This implies the following:
\begin{itemize}
\item When $n_f$ is even, $c_1(\U(n_f)) = c_1(\U(n_f)/\bZ_2)$ and $a_2(\U(n_f)/\bZ_2)=a_2(\USp(2n_f)/\bZ_2)$.
\item When $n_f$ is odd, $c_1(\U(n_f)/\bZ_2)=2 c_1(\U(n_f))$ when the latter is well-defined.
More generally, $a_2(\U(n_f)/\bZ_2)$ is the mod-2 reduction of $c_1(\U(n_f)/\bZ_2)$.
\end{itemize}
We now compute the anomaly cochains in the four cases separately:

\paragraph{$\bm{(n_c,n_f)=(\textbf{even},\textbf{even}):}$}
$w_2(\SO(2n_c))$ and $c_1(\U(n_f))$ can be generalized to closed cochains of $B(\SO(2n_c)/\bZ_2)$ 
and of $B(\U(n_f)/\bZ_2)$
without any problem,
and therefore \begin{equation}
\delta\left(\frac12 w_2(\SO(2n_c)) c_1(\U(n_f))\right) = 0.
\end{equation}

\paragraph{$\bm{(n_c,n_f)=(\textbf{odd},\textbf{even}):}$}
$w_2(\SO(2n_c))$ needs to be upgraded to a $\bZ_4$-valued cochain $x_2(\SO(2n_c)/\bZ_2)$.
The original interaction is then \begin{equation}
\frac14 x_2(\SO(2n_c)/\bZ_2) c_1(\U(n_f)),
\end{equation}
which is closed without problem,
and therefore taking $\delta$ results in zero.

\paragraph{$\bm{(n_c,n_f)=(\textbf{even},\textbf{odd}):}$}
Here we need to replace $c_1(\U(n_f))$ by $\tfrac{1}{2}c_1(\U(n_f)/\bZ_2)$.
The effective interaction is then \begin{equation}
\frac14 w_2(\SO(2n_c)) c_1(\U(n_f)/\bZ_2) 
\end{equation} and \begin{align}
\delta\left(\frac14 w_2(\SO(2n_c)) c_1(\U(n_f)/\bZ_2) \right)
&= \frac12 \left(\frac12\delta w_2(\SO(2n_c)) c_1(\U(n_f)/\bZ_2) \right) \\
&= \frac12 \Big(\beta w_2(\SO(2n_c))\Big)c_1(\U(n_f)/\bZ_2) ,
\end{align}
which is a pull-back of the anomaly cochain
\begin{equation}
 \frac12 \Big(\beta w_2(\SO(2n_c))\Big)a_2(\USp(2n_f)/\bZ_2). 
\end{equation}
This is the anomaly we wanted to see.

When $n_c=2$ and $n_f=1$, we can confirm that this is indeed the entire anomaly,
since we can compute $\mathrm{Hom}(\Omega^\text{spin}_5(B\left(\tfrac{\SO(4)\times \USp(2)}{\bZ_2}\right)),\U(1))$
and show that this is the only nontrivial element there.
For details, see Appendix~\ref{sec:so4su2}.

\paragraph{$\bm{(n_c,n_f)=(\textbf{odd},\textbf{odd}):}$}
Now we make the replacement on both sides and therefore the effective interaction is \begin{equation}
\frac18 x_2(\SO(2n_c)/\bZ_2) c_1(\U(n_f)/\bZ_2) 
\end{equation}
and \begin{align}
\delta\left(\frac18 x_2(\SO(2n_c)/\bZ_2) c_1(\U(n_f)/\bZ_2) \right)
&= \frac12 \left(\frac14\delta v_2(\SO(2n_c)/\bZ_2) c_1(\U(n_f)/\bZ_2) \right) \\
&= \frac12 \Big(\beta_2 x_2(\SO(2n_c)/\bZ_2)\Big)c_1(\U(n_f)/\bZ_2) ,
\end{align}
which is the pull-back of \begin{equation}
 \frac12 \Big(\beta_2 x_2(\SO(2n_c)/\bZ_2)\Big)a_2(\USp(2n_f)/\bZ_2).
\end{equation}
Recall that the symmetry we are now considering is $\tfrac{\SO(2n_c)\times \USp(2n_f)}{\bZ_2}$,
and therefore there is a single degree-2 obstruction cochain which equals both $v_2$ and $a_2$,
and therefore the anomaly cochain is \begin{equation}
 \frac12 x_2\beta_2 x_2.
\end{equation}
This is what we wanted to show.

\section*{Acknowledgments}
The authors thank discussions with 
Lakshya Bhardwaj,
Clay C\'ordova,
Po-Shen Hsin,
Justin Kaidi,
Ho Tat Lam,
Sakura Sch\"afer-Nameki,
Nati Seiberg,
and Yunqin Zheng.

Y.L.~is partially supported by the Programs for Leading Graduate Schools, MEXT, Japan,
via the Leading Graduate Course for Frontiers of Mathematical Sciences and Physics
and also by JSPS Research Fellowship for Young Scientists.
Y.T.~is partially supported  by JSPS KAKENHI Grant-in-Aid (Wakate-A), No.17H04837 
and also by WPI Initiative, MEXT, Japan at IPMU, the University of Tokyo.

\newpage

\appendix

\section{2-group structure, line-changing operators, and crossed module extensions}
In this paper we have encountered the 2-group extensions such as \eqref{group-extension} in massless $\mathfrak{so}$ QCD. 
Here we put our observation there into a more general framework.
A similar remark was made very recently in \cite[Sec.~2]{Bhardwaj:2021wif}.

\label{sec:remark}

\subsection{Physics setup}
\label{sec:ps}
Let us generally consider  a theory with a 0-form symmetry $G$ and a discrete 1-form symmetry $A$.
The Pontrjagin dual of the 1-form symmetry group $A$ can be identified with the following group:\footnote{
There can be nontrivial $p$-form symmetries that act trivially on all of the $p$-dimensional objects in the theory. One of the examples is the 0-form symmetries of a 3d Chern-Simons TQFT. Another example for $\mathbb{Z}_2$ 1-form symmetry is found in \cite{Hsin:2019fhf}. Such symmetries (in general topological operators) are called the \textit{condensations} \cite{Gaiotto:2019xmp}. Here we ignore these symmetries.}
\begin{equation}
    \hat A = \left\{\text{line operators}\right\}/\sim,
\end{equation}
where the quotient via $\sim$ means that we identify two line operators $L_1$ and $L_2$ if there exists a line-changing operator between them.\footnote{%
To be precise, we identify $L_1$ and $L_2$ if there exists a line operator $L_3$ such that there exists a point operator connecting $L_1, L_2^*, L_3, L_3^*$ with ${}^*$ being the orientation reversal. 
The freedom to include $L_3$ is necessary to make $A$ a group in general, for example in a 3d TQFT, but can be ignored in our non-topological gauge theory example.}

Two line operators can be connected by a line-changing operator, but the operator is not necessarily consistently acted on by the 0-form symmetry group $G$, 
which is defined to act faithfully on the local operators.
In this situation, we can also define the group
\begin{equation}
    \hat A' = \left\{\text{line operators}\right\}/\sim',
\end{equation}
where the quotient by $\sim'$ is similar to the previous one by $\sim$, but here only the line-changing operator consistently acted on by the 0-form symmetry group $G$ is considered.

This group $\hat A'$ fits in the following short exact sequence \begin{equation}
0\to \hat C\to \hat A' \to \hat A \to 0,
\end{equation} which dually forms the short exact sequence \begin{equation}
0\to A\to A'\to C\to 0. \label{coeff}
\end{equation}
The lines in $\hat C$ are equivalent to trivial lines under the equivalence relation $\sim$.
Therefore,  a line labeled by $\hat c\in \hat C$ can end on a point operator which is in a nontrivial projective representation of $G$, and $\hat c$ controls the projective phase. 
Equivalently, such a point operator is in a representation of $\tilde G$ which is an extension of $G$ by $C$: \begin{equation}
0\to C\to \tilde G\to G \to 0.\label{ext}
\end{equation}
Combining, we have an exact sequence of groups
\begin{equation}
    1 \to  A \to  A' \to \tilde{G} \to G \to 1,
\end{equation}
where $\tilde{G}$ is the group faithfully acting on the whole set of line-changing operators.
Now, the extension \eqref{ext} is characterized by an element $w_2\in H^2(G,C)$.
We can then use the Bockstein operator $\beta$ associated to \eqref{coeff} to obtain an element $\beta w_2\in H^3(G,A)$, which is the data characterizing the 2-group extension.

\subsection{Mathematical remark}
Since the dawn of time, humans wondered how  to find an interpretation for $H^3(G,A)$ and higher cohomology groups analogous to the fact that $H^2(G,A)$ classifies extensions \begin{equation}
0\to A\to \tilde G \to G \to 0.
\end{equation}
This was achieved e.g.~in \cite{Holt}.\footnote{%
It was found independently by many authors around the same time, not all of which were published.
For historical details, see \cite{MacLaneHistorical}.
}
The statement goes as follows. Given $G$ and $A$, one considers all extensions of the form \begin{equation}
0\to A\to N \stackrel{a}{\to} \tilde G \to G \to 0,
\label{cme1}
\end{equation}
where $N$ is not necessarily Abelian, 
and we furthermore require that $N$ is a \emph{crossed module over} $\tilde G$,
i.e.~there is an action of $g\in \tilde G$ on $n\in N$ which we denote as $^gn$, such that \begin{equation}
^{a(n)}n'=nn'n^{-1},\qquad a({}^gn) = ga(n) g^{-1}.
\label{cme2}
\end{equation} 
Let us denote such an extension by $(N,\tilde G)$.
For two such extensions we denote by $(N,\tilde G)\Rightarrow (N',\tilde G')$ 
if we can make the following diagram commute: \begin{equation}
\begin{array}{cccccccccc}
0\to & A &\to & N  & \to &  \tilde G & \to & G & \to 0\\
& \downarrow & &\downarrow & &\downarrow && \downarrow\\
0\to & A &\to & N'  & \to &  \tilde G' & \to & G & \to 0
\end{array}
\end{equation}
where the first and the fourth down arrows are isomorphisms and the second and the third are homomorphisms. 
Then, we say $(N,\tilde G)\approx (N',\tilde G')$ when there is a chain \begin{equation}
(N,\tilde G) \Rightarrow (N_1,\tilde G_1) \Leftarrow (N_2,\tilde G_2) \Rightarrow \cdots 
\Leftrightarrow (N',\tilde G')
\end{equation}
where the last arrow can be oriented in either direction.
The fundamental result proved in \cite{Holt} is that the extensions of the form \eqref{cme1}
satisfying \eqref{cme2} under the equivalence relation $\approx$  form the group $H^3(G,A)$.
It was further shown in \cite[Prop.~2.7]{Holt} that we can always choose $N$ to be Abelian.
In this case, the conditions \eqref{cme2} reduce to the fact that $A$ and $A'$ are $G$-modules 
and the sequence \eqref{cme1} is compatible with the $G$ action.

Therefore, our setup in Sec.~\ref{sec:ps} actually covers all possibilities of extension classes $\alpha\in H^3(G,A)$.
In particular, there always is a choice of a coefficient sequence $0\to A\to A'\to C\to 0$ \eqref{coeff} such that 
$\alpha=\beta w$ for an element $w\in H^2(G,C)$ with $\beta$ the Bockstein operation.

%

\section{Bordism group computations}
\label{sec:bordism}

The bordism groups $\Omega_\bullet^{\text{spin}}(X)$ for $X=B^{p+1}G$ are known \cite{Freed:2016rqq, Yamashita:2021cao}
to capture the  anomalies of $p$-form symmetry $G$.
More precisely, the anomalies of $d$-dimensional spin QFT are characterized by
$(d+1)$-dimensional spin invertible QFTs, whose deformation classes form a group $\mathrm{Inv}_{\text{spin}}^{d+1}(X)$
which sits in the middle of the following short exact sequence
\begin{equation}
	0
	\longrightarrow
	\mathrm{Ext}_\bZ(\Omega^{\text{spin}}_{d+1}(X),\bZ)
	\longrightarrow
	\mathrm{Inv}_{\text{spin}}^{d+1}(X)
	\longrightarrow
	\mathrm{Hom}_\bZ(\Omega^{\text{spin}}_{d+2}(X),\bZ)
	\longrightarrow
	0.
\end{equation}
Note that the information on global (non-perturbative) anomalies is encoded in the part
\begin{equation}
	\mathrm{Ext}_\bZ(\Omega^{\text{spin}}_{d+1}(X),\bZ)
	\simeq
	\mathrm{Hom}(\Omega^{\text{spin}}_{d+1}(X)_{\text{torsion}},\U(1)),
\end{equation}
while that on local (perturbative) anomalies is encoded in the part
\begin{equation}
	\mathrm{Hom}_\bZ(\Omega^{\text{spin}}_{d+2}(X),\bZ),
\end{equation}
both of which correspond to bordism invariants.

In this appendix, we compute these bordism groups $\Omega_\bullet^{\text{spin}}(X)$ for various classifying spaces,
using the Atiyah-Hirzebruch spectral sequence associated with the trivial fibration
\begin{equation*}
	pt \longrightarrow X \overset{p}{\longrightarrow} X.
\end{equation*}
In short, the spectral sequences have the $E^2$-terms given by
ordinary homology groups $H_p\big(X;\Omega_q^{\text{spin}}\big)$,
and they converge to the desired bordism groups.
For a more detailed introduction especially aimed at physicists,
see e.g.~\cite{Garcia-Etxebarria:2018ajm} and references therein.

\subsection{$X=B^2(\bZ_2\times \bZ_2)$}
The (reduced) bordism group $\tilde \Omega_d^{\text{spin}}(X)$ to be computed
characterizes the  anomalies of $\bZ_2\times \bZ_2$ 1-form symmetry in spin QFTs.
Since $B^2(\bZ_2\times \bZ_2) = B^2\bZ_2 \times B^2\bZ_2$,
the necessary information on (co)homology is derived from those of the Eilenberg-MacLane space $B^2\bZ_2 = K(\bZ_2, 2)$.
Here, the $\bZ_2$-(co)homology is known \cite{Serre1953} to be
\begin{equation}
	H^\ast(K(\bZ_2,2);\bZ_2)
	=
	\bZ_2[u_2, Sq^1 u_2, Sq^2Sq^1 u_2, \cdots],
\end{equation}
where $Sq^i$ are the Steenrod operations, among which $Sq^1$ coincides with the Bockstein homomorphism $\beta$
associated with the short exact sequence
\begin{equation}
	0
	\longrightarrow
	\bZ_2
	\longrightarrow
	\bZ_4
	\longrightarrow
	\bZ_2
	\longrightarrow
	0,
\end{equation}
while the $\bZ$-homology of $K(\bZ_2, 2)$ can be read off from \cite{Clement2002}.
Then, with the help of the K\"unneth formula which says that,
for a principal ideal domain (PID) $R$, there are short exact sequences
\begin{multline}
	\label{Kunneth}
	0
	\longrightarrow
	\bigoplus_i H_i(X;R) \otimes_R H_{n-i}(Y;R) \\[-3mm]
	\longrightarrow
	H_n(X\times Y;R)
	\longrightarrow \\[1mm]
	\bigoplus_i \mathrm{Tor}_R \big(H_i(X;R), H_{n-i-1}(Y;R)\big)
	\longrightarrow
	0
\end{multline}
which are split,
the $E^2$-page of the Atiyah-Hirzebruch spectral sequence is filled as
\begin{equation}
	\label{AHSS-K(Z2*Z2,2)}
	\begin{array}{ccc}
		E^2_{p,q}=H_p\big(K(\bZ_2\times \bZ_2,2);\Omega_q^{\text{spin}}\big)
		&& \tilde\Omega_{p+q}^{\text{spin}}(K(\bZ_2\times \bZ_2,2))\vspace{4mm}\\
		\begin{array}{c|c:cccccccccccc}
			6  &&&&&& \\
			5  & \cellcolor{lightyellow} & \hphantom{\bZ_2} & \hphantom{\bZ_2} & \hphantom{\bZ_2} & \hphantom{\bZ_2} & \hphantom{\bZ_2} \\
			4  & \bZ & \cellcolor{lightyellow} & \ast && \ast & \ast & \ast\\
			3  &  && \cellcolor{lightyellow} &&&\\
			2  & \bZ_2 &  & \red{\dbox{\black{$\bZ_2^{\oplus 2}$}}} & \Blue{\dbox{\black{$\bZ_2^{\oplus 2}$}}}\cellcolor{lightyellow} & \ast & \ast & \ast\\
			1  & \bZ_2 && \red{\fbox{\black{$\bZ_2^{\oplus 2}$}}} & \Blue{\fbox{\black{$\bZ_2^{\oplus 2}$}}} & \green{\fbox{\red{\dbox{\black{$\bZ_2^{\oplus 3}$}}}}}\cellcolor{lightyellow} & \Blue{\dbox{\black{$\bZ_2^{\oplus 6}$}}} & \ast\\
			0 & \bZ &  & \bZ_2^{\oplus 2} &  & \red{\fbox{\black{$\bZ_4^{\oplus 2}\oplus \bZ_2$}}} & \Blue{\fbox{\black{$\bZ_2^{\oplus 3}$}}} \cellcolor{lightyellow} & \green{\fbox{\black{$\ast$}}}\\
			\hline
			& 0 & 1 & 2 & 3 & 4 & 5 & 6 \\
		\end{array}
		& \quad\longrightarrow & 
		\begin{array}{c|c}
			6  & \ast\\
			5  & \bZ_2\cellcolor{lightyellow}\\
			4  & \bZ_2^{\oplus 3}\\
			3  & \\
			2  & \bZ_2^{\oplus 2}\\
			1  & \\
			0 & \\
			\hline\\
		\end{array}
	\end{array}
\end{equation}
The horizontal and vertical axes correspond to $p$ and $q$ respectively;
this will be the convention throughout the appendix.

Here, the differentials $d^2: E^2_{p,q} \to E^2_{p-2,q+1}$ for $q=0,1$
are known \cite{Teichner1993} to be the duals of $Sq^2$ (composed with mod-2 reduction for $q=0$).
First, \red{\fbox{\black{$d^2 : E^2_{4,0} \to E^2_{2,1}$}}} and
\red{\dbox{\black{$d^2 : E^2_{4,1} \to E^2_{2,2}$}}} should be duals of
\begin{equation}
	Sq^2(u_2) = (u_2)^2
\end{equation}
and also
\Blue{\fbox{\black{$d^2 : E^2_{5,0} \to E^2_{3,1}$}}} and
\Blue{\dbox{\black{$d^2 : E^2_{5,1} \to E^2_{3,2}$}}} should be duals of
\begin{equation}
	Sq^2(Sq^1 u_2) = Sq^2 Sq^1 u_2
\end{equation}
and finally \green{\fbox{\black{$d^2 : E^2_{6,0} \to E^2_{4,1}$}}} should be a dual of
\begin{equation}
	Sq^2 (u_2u'_2) = (Sq^1 u_2)(Sq^1 u'_2).
\end{equation}
As a result, the spectral sequence converges as in the RHS of \eqref{AHSS-K(Z2*Z2,2)},
and the corresponding bordism invariants in 4d are
\begin{equation}
	\dfrac{1}{2} \fP(a), \quad 
	\dfrac{1}{2} \fP(b), \quad
	a b,
\end{equation}
where $a$ (resp.~$b$) is pulled back from $u_2$ (resp.~$u'_2$),
and $\fP: H^2(-;\bZ_2)\to H^4(-;\bZ_4)$ is the Pontrjagin square.
It is known that $\fP(u_2)$ is the generator of $H^4(K(\bZ_2,2);\bZ_4)$,
and is even on a spin manifold i.e. $\fP(x) = x^2 = 0$ $(\mathrm{mod}\,2)$ for the pulled-back $x$,
which allows us to divide it by $2$.
Also, the bordism invariant in 5d is
\begin{equation}
	a\beta b
	\ 
	(= b\beta a).
\end{equation}

\newpage

\subsection{$X=B^2\bZ_4$}
This time, the bordism group to be computed
captures the  anomalies of $\bZ_4$ 1-form symmetry of spin QFTs.
It is known \cite{Serre1953} that the $\bZ_2$-cohomology ring of $B^2\bZ_4=K(\bZ_4,2)$ is
\begin{equation}
	H^\ast(K(\bZ_4,2);\bZ_2) = \bZ_2[u_2, \beta_2 \overline u_2, Sq^2\beta_2 \overline u_2, \ldots]
\end{equation}
where
$\overline u_2 \in H^2(K(\bZ_4,2);\bZ_4)$ is the $\bZ_4$-lift of $u_2$,
and $\beta_2: H^{\bullet}(-;\bZ_4) \to H^{\bullet + 1}(-;\bZ_2)$ is
the higher Bockstein operator (see also footnote \ref{foot:bock}) associated with the short exact sequence
\begin{equation*}
	0
	\longrightarrow
	\bZ_2
	\longrightarrow
	\bZ_8
	\longrightarrow
	\bZ_4
	\longrightarrow
	0.
\end{equation*}
Together with the information on the $\bZ$-homology \cite{Clement2002}, one can fill in the $E^2$-page as
\begin{equation}
	\label{AHSS-K(Z4,2)}
	\begin{array}{ccc}
		E^2_{p,q}=H_p\big(K(\bZ_4,2);\Omega_q^{\text{spin}}\big) && \tilde\Omega_{p+q}^{\text{spin}}(K(\bZ_4,2))\vspace{4mm}\\
		\begin{array}{c|c:cccccccccccc}
			6  &&&&&& \\
			5  & \cellcolor{lightyellow} & \hphantom{\bZ_2} & \hphantom{\bZ_2} & \hphantom{\bZ_2} & \hphantom{\bZ_2} & \hphantom{\bZ_2} \\
			4  & \bZ & \cellcolor{lightyellow} & \ast && \ast & \ast & \ast\\
			3  &  && \cellcolor{lightyellow} &&&\\
			2  & \bZ_2 &  & \red{\dbox{\black{$\bZ_2$}}} & \Blue{\dbox{\black{$\bZ_2$}}}\cellcolor{lightyellow} & \ast & \ast & \ast\\
			1  & \bZ_2 && \red{\fbox{\black{$\bZ_2$}}} & \Blue{\fbox{\black{$\bZ_2$}}} & \red{\dbox{\black{$\bZ_2$}}}\cellcolor{lightyellow} & \Blue{\dbox{\black{$\ast$}}} & \ast\\
			0 & \bZ &  & \bZ_4 &  & \red{\fbox{\black{$\bZ_8$}}} & \Blue{\fbox{\black{$\bZ_2$}}} \cellcolor{lightyellow} & \ast\\
			\hline
			& 0 & 1 & 2 & 3 & 4 & 5 & 6 \\
		\end{array}
		& \quad\longrightarrow & 
		\begin{array}{c|c}
			6  & \ast\\
			5  & \cellcolor{lightyellow}\\
			4  & \bZ_4\\
			3  & \\
			2  & \bZ_4\\
			1  & \\
			0 & \\
			\hline\\
		\end{array}
	\end{array}
\end{equation}
As before, the differentials
\red{\fbox{\black{$d^2 : E^2_{4,0} \to E^2_{2,1}$}}} and
\red{\dbox{\black{$d^2 : E^2_{4,1} \to E^2_{2,2}$}}} should be duals of
\begin{equation}
	Sq^2(u_2) = (u_2)^2
\end{equation}
while
\Blue{\fbox{\black{$d^2 : E^2_{5,0} \to E^2_{3,1}$}}} and
\Blue{\dbox{\black{$d^2 : E^2_{5,1} \to E^2_{3,2}$}}} should be duals of
\begin{equation}
	Sq^2(\beta_2 \overline u_2) = Sq^2 \beta_2 \overline u_2.
\end{equation}
Therefore, the spectral sequence converges as in the RHS of \eqref{AHSS-K(Z4,2)},
and the bordism invariant in 4d is simply given by (multiples of)
\begin{equation}
		\frac12\fP(a) 
\end{equation}
where $a$ is pulled back from $\overline u_2$,
and $\fP: H^2(-;\bZ_4)\to H^4(-;\bZ_8)$ is the Pontrjagin square,
which is again even on a spin manifold and thus divisible by $2$.
In contrast, there are no bordism invariants in 5d.

\newpage

\subsection{$X=B\left(\tfrac{\SO(4)\times \SU(2)}{\bZ_2}\right)$}
\label{sec:so4su2}

The necessary information on (co)homology can be obtained
by using the Leray-Serre spectral sequence,
whose $E_2$-terms are $H^p(B; H^q(F;\bZ))$ and converges to $H^{\bullet}(E;\bZ)$
for the fibration
	$F
	\longrightarrow
	E
	\overset{p}{\longrightarrow}
	B$.
For the case of interest, the relevant fibration is
\begin{equation}
	B\SU(2)
	\longrightarrow
	B\left(\tfrac{\SO(4)\times \SU(2)}{\bZ_2}\right)
	\longrightarrow
	B\left(\SO(4)/\bZ_2\right)
	=
	B\SO(3)\times B\SO(3)
\end{equation}
where the cohomology of the fiber is known to be
\begin{equation}
	H^\ast(B\SU(2);\bZ) = \bZ[c_2]
\end{equation}
while that of the base is derived from
\begin{equation}
	H^\ast(B\SO(3);\bZ_2) = \bZ_2[w_2, w_3]
\end{equation}
and
\begin{equation}
	\begin{array}{c||cccccccc}
		d & 0 & 1 & 2 & 3 & 4 & 5 & 6 & \cdots\\
		\hline
		H^d(B\SO(3);\bZ) & \bZ & 0 & 0 & \bZ_2 & \bZ & 0 & \bZ_2 & \cdots\\
	\end{array}
\end{equation}
together with the use of the K\"unneth formula.
As a result, the $E_2$-page is filled as
\begin{equation}
	\begin{array}{ccc}
		E_2^{p,q}=H^p\big(B\left(\SO(4)/\bZ_2\right);H^q(B\SU(2);\bZ)\big) && H^{p+q}(B\left(\tfrac{\SO(4)\times \SU(2)}{\bZ_2}\right);\bZ)\vspace{4mm}\\
		\begin{array}{c|ccccccccccccc}
			6  &&&&&& \\
			5  & \cellcolor{lightyellow} & \hphantom{\bZ_2} & \hphantom{\bZ_2} & \hphantom{\bZ_2} & \hphantom{\bZ_2} & \hphantom{\bZ_2} \\
			4  & \red{\fbox{\black{$\bZ$}}} & \cellcolor{lightyellow} && \ast & \ast & \ast & \ast\\
			3  &  && \cellcolor{lightyellow} &&&\\
			2  &&  && \cellcolor{lightyellow} &&\\
			1  &  &&  && \cellcolor{lightyellow} &&\\
			0 & \bZ &  && \bZ_2^{\oplus 2} & \bZ^{\oplus 2} & \red{\fbox{\black{$\bZ_2$}}}\cellcolor{lightyellow} & \bZ_2^{\oplus 3}\\
			\hline
			& 0 & 1 & 2 & 3 & 4 & 5 & 6 \\
		\end{array}
		& \longrightarrow & 
		\begin{array}{c|c}
			6  & \bZ_2^{\oplus 3}\\
			5  & \cellcolor{lightyellow}\\
			4  & \bZ^{\oplus 3}\\
			3  & \bZ_2^{\oplus 2}\\
			2  & \\
			1  & \\
			0 & \bZ\\
			\hline\\
		\end{array}
	\end{array}
\end{equation}
It turns out that
the differential \red{\fbox{\black{$d_5 : E_{0,4} \to E_{5,0}$}}} must be nontrivial
to account for the allowed instanton numbers.\footnote{%
To explain this point in more detail, 
note first that the $E_2$-page implies that $H^4(B\left(\tfrac{\SO(4)\times\SU(2)}{\bZ_2}\right);\bZ)=\bZ^{\oplus3}$
regardless of whether the differential is trivial.
Recalling that $\tfrac{\SO(4)\times \SU(2)}{\bZ_2}$ contains three $\mathfrak{su}(2)$ factors,
let $c_2$, $c_2'$ be the instanton numbers of two $\mathfrak{su}(2)$ factors of the $\SO(4)/\bZ_2$ part,
so that $p_1=4c_2$ and $p_1'=4c_2'$ are the generators of $E_{4,0}$. 
Similarly, let $\tilde c_2$ be the instanton number of the $\SU(2)$ part, i.e.~the generator of $E_{0,4}$.
Now, in $B\left(\tfrac{\SO(4)\times\SU(2)}{\bZ_2}\right)$,  
we have $(c_2,c_2',\tilde c_2) = \tfrac14 (\fP(w_2+a_2),\fP(w_2),\fP(a_2))$ modulo $\bZ^3$;
this simply follows from the fact that $p_1=\fP(w_2)$ mod 4 in $B\SO(3)$ \cite{Thomas}.
Then, we see that $2(c_2+c_2'+\tilde c_2)$ is always $\bZ$-valued, meaning that $H^4(B\left(\tfrac{\SO(4)\times\SU(2)}{\bZ_2}\right);\bZ)=\bZ^{\oplus3}$ is obtained by extending
$H^4(B(\SO(4)/\bZ_2);\bZ)=\bZ^{\oplus2}$ by the $\bZ$ generated by $2\tilde c_2$.
This means that the differential $d_5$ in question needs to be a mod-2 reduction i.e. nontrivial.}
As a result, we end up with the following integral cohomology structure
\begin{equation}
	\label{case1}
	\renewcommand{\arraystretch}{1.2}
	\begin{array}{c||cccccccccccccccc}
		d & 0 & 1 & 2 & 3 & 4 & 5 & 6 & \cdots \\
		\hline
		H^d(B\left(\tfrac{\SO(4)\times \SU(2)}{\bZ_2}\right);\bZ) & \bZ & 0 & 0 & \bZ_2^{\oplus 2} & \bZ^{\oplus 3} & 0 & \bZ_2^{\oplus 3} & \cdots\\
	\end{array}
\end{equation}

Having obtained the (co)homology groups,
one can fill in the $E^2$-page of the Atiyah-Hirzebruch spectral sequence:
\begin{equation}
	\begin{array}{ccc}
		E^2_{p,q}=H_p\big(B\left(\tfrac{\SO(4)\times \SU(2)}{\bZ_2}\right);\Omega_q^{\text{spin}}\big)\vspace{2mm}\\
		\begin{array}{c|c:cccccccccccc}
			6  &&&&&& \\
			5  & \cellcolor{lightyellow} & \hphantom{\bZ_2} & \hphantom{\bZ_2} & \hphantom{\bZ_2} & \hphantom{\bZ_2} & \hphantom{\bZ_2} \\
			4  & \bZ & \cellcolor{lightyellow} & \bZ_2^{\oplus 2} && \ast & \ast & \ast\\
			3  &  && \cellcolor{lightyellow} &&&\\
			2  & \bZ_2 &  & \red{\dbox{\black{$\bZ_2^{\oplus 2}$}}} & \Blue{\dbox{\black{$\bZ_2^{\oplus 2}$}}}\cellcolor{lightyellow} & \bZ_2^{\oplus 3} & \ast & \ast\\
			1  & \bZ_2 && \red{\fbox{\black{$\bZ_2^{\oplus 2}$}}} & \Blue{\fbox{\black{$\bZ_2^{\oplus 2}$}}} & \green{\fbox{\black{\red{\dbox{\black{$\bZ_2^{\oplus 3}$}}}}}}\cellcolor{lightyellow} & \Blue{\dbox{\black{$\bZ_2^{\oplus 3}$}}} & \ast\\
			0 & \bZ &  & \bZ_2^{\oplus 2} &  & \red{\fbox{\black{$\bZ^{\oplus 3}$}}} & \Blue{\fbox{\black{$\bZ_2^{\oplus 3}$}}} \cellcolor{lightyellow} & \green{\fbox{\black{$\ast$}}}\\
			\hline
			& 0 & 1 & 2 & 3 & 4 & 5 & 6 \\
		\end{array}
	\end{array}
\end{equation}
For each differential,
\red{\fbox{\black{$d^2 : E^2_{4,0} \to E^2_{2,1}$}}} and
\red{\dbox{\black{$d^2 : E^2_{4,1} \to E^2_{2,2}$}}} should be duals of
\begin{equation}
	Sq^2(w_2) = (w_2)^2
\end{equation}
and also
\Blue{\fbox{\black{$d^2 : E^2_{5,0} \to E^2_{3,1}$}}} and
\Blue{\dbox{\black{$d^2 : E^2_{5,1} \to E^2_{3,2}$}}} should be duals of
\begin{equation}
	Sq^2(w_3) = w_2w_3
\end{equation}
and finally \green{\fbox{\black{$d^2 : E^2_{6,0} \to E^2_{4,1}$}}} should be a dual of
\begin{equation}
	Sq^2 (w_2w'_2) = w_3w'_3 + (w_2)^2w'_2 + w_2(w'_2)^2.
\end{equation}
Then, the $E^3$-page would become
\begin{equation}
	\label{e3page1}
	\begin{array}{ccc}
		E^3_{p,q} && \tilde\Omega_{p+q}^{\text{spin}}(
			B\left(\tfrac{\SO(4)\times \SU(2)}{\bZ_2}\right)
		)\vspace{2mm}\\
		\begin{array}{c|c:cccccccccccc}
			6  &&&&&& \\
			5  & \cellcolor{lightyellow} & \hphantom{\bZ_2} & \hphantom{\bZ_2} & \hphantom{\bZ_2} & \hphantom{\bZ_2} & \hphantom{\bZ_2} \\
			4  & \bZ & \cellcolor{lightyellow} & \ast && \ast & \ast & \ast\\
			3  &  && \cellcolor{lightyellow} &&&& \hphantom{\bZ_2}\\
			2  & \bZ_2 &  && \cellcolor{lightyellow} & \ast & \ast & \ast\\
			1  & \bZ_2 &&  && \cellcolor{lightyellow} & \ast & \ast\\
			0 & \bZ &  & \bZ_2^{\oplus 2} &  & \bZ^{\oplus 3} & \bZ_2\cellcolor{lightyellow} & \ast\\
			\hline
			& 0 & 1 & 2 & 3 & 4 & 5 & 6 \\
		\end{array}
		& \quad\longrightarrow & 
		\begin{array}{c|c}
			6  & \ast\\
			5  & \bZ_2\cellcolor{lightyellow}\\
			4  & \bZ^{\oplus 3}\\
			3  & \\
			2  & \bZ_2^{\oplus 2}\\
			1  & \\
			0 & \\
			\hline\\
		\end{array}
	\end{array}
\end{equation}
and converges to the RHS.
Therefore, the bordism invariant in 5d characterizing the anomaly of interest is
\begin{equation}
	a \beta b
	\ 
	(= b \beta a)
\end{equation}
where $a$ (resp.~$b$) is pulled back from $w_2$ (resp.~$w'_2$).

\subsection{$X=B\left(\tfrac{\SO(4n+2)\times \SU(2m)}{\bZ_2}\right)$}
\label{sec:so(2odd)}

We again use the Leray-Serre spectral sequence, this time for the fibration
\begin{equation}
	B\SU(2m)
	\longrightarrow
	B\left(\tfrac{\SO(4n+2)\times \SU(2m)}{\bZ_2}\right)
	\longrightarrow
	B\left(\tfrac{\SO(4n+2)}{\bZ_2}\right)
	=
	B\mathrm{PSO}(4n+2).
\end{equation}
According to \cite{KonoMimura1974},
the $\bZ_2$-cohomology of $B\mathrm{PSO}$ is given as follows
\begin{equation}
	\label{BPSO-Z2coho}
	\renewcommand{\arraystretch}{1.2}
	\begin{array}{c||cccccccccccccccc}
		d & 0 & 1 & 2 & 3 & 4 & 5 & 6 & \cdots \\
		\hline
		\multicolumn{1}{l||}{H^d(B\mathrm{PSO}(4n+2);\bZ_2)} & \bZ_2 & 0 & \bZ_2 & \bZ_2 & \bZ_2 & \bZ_2 & \bZ_2^{\oplus 2} & \cdots\\
		\hline
		\text{generators} & 1 & - & v_2 & y'(1) & (v_2)^2 & y'(2) & (v_2)^3 & \cdots\\
		&&&&&&& y'(1)^2\\
		\hline
	\end{array}
\end{equation}
for $n\geq 1$, where the action of the cohomology operations are
\begin{equation}
	\begin{array}{ccl}
		\beta_2 \overline v_2 & = & y'(1),\\
		Sq^2 y'(1) & = & y'(2),\\
		Sq^1 y'(2) & = & y'(1)^2.\\
	\end{array}
\end{equation}
The $\bZ$-cohomology of $B\mathrm{PSO}$ can be determined
by exploiting another Leray-Serre spectral sequence for the fibration
\begin{equation}
	B\SO(4n+2)
	\longrightarrow
	B\mathrm{PSO}(4n+2)
	\longrightarrow
	B^2\bZ_2
	=
	K(\bZ_2, 2).
\end{equation}
From the knowledge on the $\bZ$-cohomology of
$B\SO(4n+2)$ \cite{Brown1982, Feschbach1983} and
$K(\bZ_2,2)$ \cite{Clement2002},
the $E_2$-page is filled as
\begin{equation}
	\begin{array}{c}
		E_2^{p,q}=H^p\big(B^2\bZ_2;H^q(B\SO(4n+2);\bZ)\big)\vspace{4mm}\\
		\begin{array}{c|ccccccc}
			6 & \bZ_2 & \hphantom{\bZ_2} & \ast & \ast & \ast & \ast & \ast\\
			5 & \bZ_2 \cellcolor{lightyellow}&& \ast & \ast & \ast & \ast &\ast \\
			4 & \bZ & \cellcolor{lightyellow} && \ast & \hphantom{\bZ_2} & \ast & \ast\\
			3 & \bZ_2\cellcolor{lightyellow}  && \bZ_2\cellcolor{lightyellow} & \bZ_2 & \ast & \ast & \ast\\
			2 && \cellcolor{lightyellow} && \cellcolor{lightyellow} && \\
			1 &  && \cellcolor{lightyellow} && \cellcolor{lightyellow} &\\
			0 & \bZ &  && \bZ_2\cellcolor{lightyellow}  && \bZ_4 \cellcolor{lightyellow} & \bZ_2\\
			\cline{2-8}
			& 0 & 1 & 2 & 3 & 4 & 5 & 6
		\end{array}
	\end{array}
\end{equation}
from which one can deduce
\begin{equation}
	\renewcommand{\arraystretch}{1.2}
	\begin{array}{c||cccccccccccccccc}
		d & 0 & 1 & 2 & 3 & 4 & 5 & 6 & \cdots \\
		\hline
		\multicolumn{1}{l||}{H^d(B\mathrm{PSO}(4n+2);\bZ)} & \bZ & 0 & 0 & ?\cellcolor{lightyellow} & \bZ & ?\cellcolor{lightyellow} & ? & \cdots
	\end{array},
\end{equation}
and for example the $d=3$ piece is either $\bZ_2\times \bZ_2$ or $\bZ_4$.
By requiring the result to be consistent with the $\bZ_2$-cohomology \eqref{BPSO-Z2coho}
and the universal coefficient theorem,
one can  actually conclude
\begin{equation}
	\renewcommand{\arraystretch}{1.2}
	\begin{array}{c||cccccccccccccccc}
		d & 0 & 1 & 2 & 3 & 4 & 5 & 6 & \cdots \\
		\hline
		\multicolumn{1}{l||}{H^d(B\mathrm{PSO}(4n+2);\bZ)} & \bZ & 0 & 0 & \bZ_4 & \bZ & 0 & \bZ_2 & \cdots
	\end{array}
\end{equation}
Then, the $E_2$-page of the original Leray-Serre spectral sequence can be filled and converges as
\begin{equation}
	\begin{array}{ccc}
		E_2^{p,q}=H^p\big(B\left(\tfrac{\SO(4n+2)}{\bZ_2}\right);H^q(B\SU(2m);\bZ)\big) && H^{p+q}(B\left(\tfrac{\SO(4n+2)\times \SU(2m)}{\bZ_2}\right);\bZ)\vspace{4mm}\\
		\begin{array}{c|ccccccccccccc}
			6  & \bZ &&& \ast & \ast && \ast\\
			5  & \cellcolor{lightyellow} & \hphantom{\bZ_2} & \hphantom{\bZ_2} & \hphantom{\bZ_2} & \hphantom{\bZ_2} & \hphantom{\bZ_2} \\
			4  & \bZ & \cellcolor{lightyellow} && \ast & \ast & & \ast\\
			3  &  && \cellcolor{lightyellow} &&&\\
			2  &&  && \cellcolor{lightyellow} &&\\
			1  &  &&  && \cellcolor{lightyellow} &&\\
			0 & \bZ &  && \bZ_4 & \bZ & \cellcolor{lightyellow} & \bZ_2\\
			\hline
			& 0 & 1 & 2 & 3 & 4 & 5 & 6 \\
		\end{array}
		& \longrightarrow & 
		\begin{array}{c|c}
			6  & \bZ\oplus \bZ_2\\
			5  & \cellcolor{lightyellow}\\
			4  & \bZ^{\oplus 2}\\
			3  & \bZ_4\\
			2  & \\
			1  & \\
			0 & \bZ\\
			\hline\\
		\end{array}
	\end{array}.
\end{equation}

Having obtained the (co)homology groups,
one can fill in the $E^2$-page of the Atiyah-Hirzebruch spectral sequence as follows:
\begin{equation}
	\label{B([SO*SU]_Z2)-generic}
	\begin{array}{ccc}
		E^2_{p,q}=H_p\big(B\left(\tfrac{\SO(4n'_c+2)\times \SU(2n_f)}{\bZ_2}\right);\Omega_q^{\text{spin}}\big)
		&& \tilde\Omega_{p+q}^{\text{spin}}(
			B\left(\tfrac{\SO(4n+2)\times \SU(2m)}{\bZ_2}\right)
		)\vspace{2mm}\\
		\begin{array}{c|c:cccccccccccc}
			6  &&&&&& \\
			5  & \cellcolor{lightyellow} & \hphantom{\bZ_2} & \hphantom{\bZ_2} & \hphantom{\bZ_2} & \hphantom{\bZ_2} & \hphantom{\bZ_2} \\
			4  & \bZ & \cellcolor{lightyellow} & \ast && \ast & \ast & \ast\\
			3  &  && \cellcolor{lightyellow} &&&\\
			2  & \bZ_2 &  & \red{\dbox{\black{$\bZ_2$}}} & \Blue{\dbox{\black{$\bZ_2$}}}\cellcolor{lightyellow} & \ast & \ast & \ast\\
			1  & \bZ_2 && \red{\fbox{\black{$\bZ_2$}}} & \Blue{\fbox{\black{$\bZ_2$}}} & \green{\fbox{\black{\red{\dbox{\black{$\bZ_2^{\oplus 2}$}}}}}}\cellcolor{lightyellow} & \Blue{\dbox{\black{$\ast$}}} & \ast\\
			0 & \bZ &  & \bZ_4 &  & \red{\fbox{\black{$\bZ^{\oplus 2}$}}} & \Blue{\fbox{\black{$\bZ_2$}}}\cellcolor{lightyellow} & \green{\fbox{\black{$\ast$}}}\\
			\hline
			& 0 & 1 & 2 & 3 & 4 & 5 & 6 \\
		\end{array}
		& \quad\longrightarrow & 
		\begin{array}{c|c}
			6  & \ast\\
			5  & \cellcolor{lightyellow}\\
			4  & \bZ^{\oplus 2}\\
			3  & \\
			2  & \bZ_4\\
			1  & \\
			0 & \\
			\hline\\
		\end{array}
	\end{array}.
\end{equation}
For each differential,
\red{\fbox{\black{$d^2 : E^2_{4,0} \to E^2_{2,1}$}}} and
\red{\dbox{\black{$d^2 : E^2_{4,1} \to E^2_{2,2}$}}} should be duals of
\begin{equation}
	Sq^2 v_2 = (v_2)^2
\end{equation}
and also
\Blue{\fbox{\black{$d^2 : E^2_{5,0} \to E^2_{3,1}$}}} and
\Blue{\dbox{\black{$d^2 : E^2_{5,1} \to E^2_{3,2}$}}} should be duals of
\begin{equation}
	Sq^2 y'(1) = y'(2)\\
\end{equation}
and finally
\green{\fbox{\black{$d^2 : E^2_{6,0} \to E^2_{4,1}$}}} should be a dual of
\begin{equation}
	Sq^2 c_2 = c_3.
\end{equation}
Therefore, the spectral sequence converges as in the RHS of \eqref{B([SO*SU]_Z2)-generic},
and in particular there should be no bordism invariants in 5d.

\newpage

\section{Coboundary of Pontrjagin square for non-closed cochains}
\label{sec:nonclosedP}
The aim of this section is to determine the coboundary of the Pontrjagin square of non-closed cochains.
Recall that the Pontrjagin square for an element $x\in C^\bullet(-;\bZ_{2^m})$
is defined to be \begin{equation}
\fP(x):= \tilde x \cup\tilde x - \tilde x \cup_1 \delta \tilde x
\end{equation} where $\tilde x \in C^\bullet(-;\bZ)$ is an integral lift of $x$,
and $\cup_1$ is the higher cup product of Steenrod.
The variation of interest is then given by 
\begin{equation}
	\label{deltaPontrjagin}
	\begin{array}{ccl}
		\delta \left(\dfrac{1}{2^{m+1}}\fP(x)\right)
		& = & \dfrac{1}{2^{m+1}}\cdot \delta \Big(\tilde x \cup \tilde x - \tilde x \cup_1 \delta \tilde x\Big)\\
		& = & \dfrac{1}{2^{m+1}}\cdot\left[\Big(\delta \tilde x \cup \tilde x + \tilde x \cup \delta \tilde x\Big)
				- \Big(\tilde x \cup \delta \tilde x - \delta \tilde x \cup \tilde x + \delta \tilde x \cup_1 \delta \tilde x\Big)\right]\vspace{2mm}\\
		& = & \dfrac{1}{2^{m+1}}\cdot\Big[2\cdot \delta \tilde x \cup \tilde x - \delta \tilde x \cup_1 \delta \tilde x\Big].\\
	\end{array}
\end{equation}
If $x$ is a $\bZ_{2^m}$-cocycle, 
then $\tilde x$ is a cocycle mod $2^m$ i.e. $\delta \tilde x = 0$ (mod $2^m$),
and the RHS of \eqref{deltaPontrjagin} is $0$ mod $1$,
which then means that $\fP(x)$ is a $\bZ_{2^{m+1}}$-cocycle. 
However, when $x$ is not a cocycle but merely a cochain,
$\fP(x)$ is also not a cocycle.
For our purpose, we limit ourselves to the case
\begin{equation}
\delta x = \beta y
\end{equation} 
for a cocycle $y\in Z^2(-;\bZ_2)$.
Recalling that the Bockstein operation $\beta$ is associated with the short exact sequence
\begin{equation}
	0
	\longrightarrow
	\bZ_2
	\stackrel{2}{\longrightarrow}
	\bZ_4
	\longrightarrow
	\bZ_2
	\longrightarrow
	0,
\end{equation}
$x$ and $y$ combine to define a cocycle $z\in Z^2(-;\bZ_4)$,
such that $z = y$ (mod $2$) and $\tilde z=2\tilde x$ when $y=0$.
This motivates us to consider the term \begin{equation}
\frac 14 \cdot \frac14 \fP(\tilde z) ,
\end{equation} which reduces to $\frac14 \fP(\tilde x)$ when $y=0$,
as its general replacement.
Using \eqref{deltaPontrjagin}, we find \begin{equation}
	\begin{array}{ccl}
		\delta\left(
			\dfrac{1}{4}\cdot\dfrac{1}{4} \fP(\tilde z)
		\right)
		& = & \dfrac{1}{2} \cdot \dfrac{1}{8} \cdot \Big[2 \cdot \delta\tilde z \cup \tilde z - \delta \tilde z \cup_1 \delta \tilde z\Big]\vspace{2mm}\\
		& = & \dfrac{1}{2} \cdot \left(\dfrac{1}{4} \delta \tilde z\right) \cup \tilde z - \left(\dfrac{1}{4} \delta \tilde z\right) \cup_1 \left(\dfrac{1}{4} \delta \tilde z\right)\vspace{2mm}\\
		& \stackrel{\mathrm{mod}\ 2}{=} & \dfrac{1}{2} \cdot (\beta_2 z) \cup z \mod 1
	\end{array}
\end{equation}
where $\beta_2$ is the higher Bockstein operation associated with the short exact sequence
\begin{equation}
	\begin{array}{ccccccccc}
		0
		& \longrightarrow & \bZ_2
		& \stackrel{4}{\longrightarrow} & \bZ_8
		& \longrightarrow & \bZ_4
		& \longrightarrow & 0\\
		&& \uparrow
		&& \uparrow
		&& \rotatebox[]{90}{$=$}\\
		0
		& \longrightarrow & \bZ
		& \stackrel{4}{\longrightarrow} & \bZ
		& \longrightarrow & \bZ_4
		& \longrightarrow & 0\\		
	\end{array}
\end{equation}
defined for cocycles $y\in Z^2(-;\bZ_2)$ which are $\bZ_4$-liftable to $z\in Z^2(-;\bZ_4)$.

\newpage

\def\arxivfont{\rm}
\bibliographystyle{ytamsalpha}

\baselineskip=.98\baselineskip
\let\originalthebibliography\thebibliography
\renewcommand\thebibliography[1]{
  \originalthebibliography{#1}
  \setlength{\itemsep}{0pt plus 0.3ex}
}

\bibliography{ref}

\end{document}